\documentclass[journal]{IEEEtran}
\usepackage{cite}
\usepackage{amsmath,amssymb,amsfonts}
\usepackage{algorithmic}
\usepackage{graphicx}
\usepackage{textcomp}
\usepackage{multirow}
\usepackage[caption=false,font=footnotesize]{subfig}
\usepackage[hyphens]{url}
\usepackage{xcolor}

\begin{document}
\title{XLR (piXel Loss Rate): a Lightweight Indicator to Measure Video QoE in IP Networks}

\author{C\'esar~D\'iaz,
        Pablo~P\'erez,
        Juli\'an~Cabrera,
        Jaime~J.~Ruiz,
        and~Narciso~Garc\'ia% <-this % stops a space
\thanks{C. D\'iaz, J. Cabrera and N. Garc\'ia are with the Grupo de Tratamiento
de Im\'agenes, Information Processing and Telecommunications Center
and ETSI Telecomunicaci\'on, Universidad Polit\'ecnica de Madrid, 28040
Madrid, Spain (e-mail: cdm@gti.ssr.upm.es; julian.cabrera@gti.ssr.upm.es;
narciso@gti.ssr.upm.es).}% <-this % stops a space
\thanks{P. P\'erez and J. J. Ruiz are with Nokia Bell Labs, Mar\'ia Tubau~9,
28050 Madrid, Spain (e-mail: pablo.perez@nokia-bell-labs.com; jaime\_jesus.ruiz\_alonso@nokia-bell-labs.com).}% <-this % stops a space
\thanks{Manuscript received XXX, 2019; revised YYY, 2019. This work has been partially supported by the Ministerio de Ciencia, Innovaci\'on y Universidades (AEI/FEDER) of the Spanish Government under projects RTC-2015-4133-7 (MOVISE) and TEC2016-75981 (IVME).}}


\markboth{
IEEE Transactions on Network and Service Management,~Vol.~XX, No.~YY, ZZ}%
{D\'iaz \MakeLowercase{\textit{et al.}}: XLR (piXel Loss Rate): a Lightweight Indicator to Measure Video QoE}



\maketitle

\begin{abstract}
A novel Key Quality Indicator for video delivery applications, XLR (piXel Loss Rate), is defined, characterized, and evaluated. The proposed indicator is an objective measure that captures the effects of transmission errors in the received video, has a good correlation with subjective Mean Opinion Scores, and provides comparable results with state-of-the-art Full-Reference metrics. Moreover, XLR can be estimated using only a lightweight analysis on the compressed bitstream, thus allowing a No-Reference operational method. Therefore, XLR can be used for measuring the quality of experience without latency at any network location. Thus, it is a relevant tool for network planning, specially in new high-demanding scenarios. The experiments carried out show the outstanding performance of its linear-dimension score and the reliability of the bitstream-based estimation.
\end{abstract}

\begin{IEEEkeywords}
Quality of experience, Video delivery, KQI, Network monitoring, IP networks, Quality measurement.
\end{IEEEkeywords}

% \fi
\IEEEpeerreviewmaketitle

\section{Introduction}
\label{sec:introduction}
\IEEEPARstart{M}{odern} IP networks are characterized by the uptake of an immense number of human- and machine-based communications in many economic sectors and vertical industries~\cite{2016_osseiran}. In this scenario, video plays a fundamental role, as it is a central part of countless specific systems and services in many areas: entertainment, education, videoconferencing, surveillance, computer vision, etc., and now also in many Internet of Things (IoT) applications. Thus, not surprisingly, video traffic accounts for more than 78\% of all IP traffic today, a percentage that is expected to continue growing over the next three years~\cite{2019-cisco}. Furthermore, in absolute numbers, IP video consumption will grow twofold over the same period. Moreover, it is also noteworthy that transmitted videos are of an increasingly higher quality. Indeed, Ultra High Definition (UHD)-4K and Full HD (FHD)-2K video combined will account for almost 80\% of all transmitted IP video~\cite{2019-cisco}. Despite the recent development and planning of new more efficient video compression formats like High Efficiency Video Coding (HEVC)~\cite{2012-sullivan}, Versatile Video Coding (VVC)~\cite{2019-tanau}, VP9~\cite{2016-grange}, or AOMedia Video~1 (AV1)~\cite{2019-aom}, these figures mean that the bandwidth consumed by video-related applications will increase globally tremendously in the following years.

Hence, IP networks should be able to provide dynamic management and deliver high-class video quality at less resources~\cite{2016_uusitalo}. In this respect, the challenges are set in the first place around supplying video to any device at any time with adequate quality of service (QoS). In this sense, there are several Key Performance Indicators (KPIs) that are relevant to monitor the service's performance, such as data rate, end-to-end latency and reliability: maximum tolerable packet loss ratio for that application~\cite{2017_5gppp}. However, meeting certain QoS goals does not necessarily result in a satisfying service from the user's point of view. Hence, service and content providers are becoming more and more interested in using user-centric Key Quality Indicators (KQIs) to monitor their solutions~\cite{2013-devi}. These KQIs are used to measure the impact of system internal and external dynamics in the users' quality of experience (QoE)~\cite{2018-li}.

In real-time video streaming scenarios, which impose high data rate and low latency requirements, the management of lost packets is key to provide an adequate performance. In this context, although widely used today to provide high quality video~\cite{2019-seufert, 2019-kumar}, TCP's inherent mechanisms to guarantee reliable communications (i.e. flow, congestion, and error control) complicate the provision of data at a sufficiently high rate and low latency, thus making it very difficult to fulfill the requirements under consideration. Indeed, even recently released formats focused on providing video with low latency over TCP (e.g. MPEG's Common Media Application Format -CMAF-~\cite{2018-cmaf}) introduce an overall delay in the order of seconds, comparable to that of linear broadcasts, which is still unacceptable for many applications and services. In these situations, conversational protocols in the application layer are typically based on RTP/UDP, always accepting that eventually there will be video packets not arriving on time at the video decoder~\cite{2011-perez}. This is the case of a great number of services and applications from the ones integrated in operational IPTV ecosystems to the strategies based on Web-based Real-Time Communication (WebRTC)~\cite{2014-rhinow, 2017-ha}, and of new network infrastructures, like Vehicular Ad-Hoc Networks (VANETs)~\cite{2016-mammeri} or IoT networks~\cite{2017-said}. In these cases, the KQIs are typically obtained from the statistics provided by the peers. Although there are several indicators that can be checked, they end up being reduced to the standard QoS scores~\cite{2013-fund, 2016-ammar}: bit rate, latency-related measurements (e.g. delay, or jitter), and packet loss rate (PLR)~\cite{2016_mustag,2017-tsolkas}. These indicators might be used as building blocks for more complex scores based on machine learning~\cite{2018-sulema, 2019-tao}.

Thus, implicitly, PLR is widely used not only as a KPI, but also as a KQI of the presented video. Although the main reason behind it is simplicity, it is not a bad decision for the long term, as packet loss rate is known to be a good average predictor of media quality in this time horizon~\cite{2004-reibman,2010_fiedler}. However, not all the video packets have the same impact when lost, due to the spatiotemporal prediction structures of video coding systems~\cite{2011_perez_b}. Under packet losses, due to error propagation through interdependent data, there is a domino effect that affects the video quality (either video freezing or 'slicing', depending on the error concealment model) from the packet loss until video data is refreshed, typically through a refresh frame~\cite{2013_garcia}. This fact can be used, for instance, to perform unequal packet protection~\cite{2011-perez,2017-diaz} or selective packet discard~\cite{2005-Cisco} depending on the impact of losing the packet. As PLR does not consider any differences between packets, it is insufficient to properly assess the actual quality of the images presented to users and so it is not suited as a KQI in the short and medium terms.

There exist a number of metrics that are suited to some extent for QoE measurement under coding and transmission errors: Mean Squared Error (MSE), Peak Signal-to-Noise Ratio (PSNR), Structural Similarity (SSIM) index, Multi-scale SSIM (MS-SSIM), Video Quality Metric (VQM), etc~\cite{2018-yang}. However, they are Full-Reference (FR) methods that cannot be estimated through No-Reference (NR) approaches. Hence, several works have developed bitstream-based QoE metrics that can predict the user's mean opinion score (MOS) based on objective characteristics of the video~\cite{2017-zhao}, many looking for international standardization~\cite{2012_itu,2012_itu_b}. There are two main types of approaches. Some of them estimate the users' opinion by means of a parametric fitting of subjective assessment tests~\cite{2012_yang,2016_juluri}. The others are based on accessing and inspecting in depth the bitstream to obtain coding-related outcomes throughout the sequence, like motion vectors and residuals, that allow to infer to some extent some local and global features of the sequence content~\cite{2012_lin, 2017_chen, 2018-panetta}. However, neither type has succeeded to be widely used as KQIs for network planning or system inter-operability. We can imagine some reasons about it. On the one hand, even though they are 'objective' in the sense that they are measured by an automatic system, and the association is done with enough care to be able to generalize, it is often difficult to extrapolate it to external values when conditions change. On the other hand, even though those metrics are not extremely costly in terms of resource consumption, they still require either measuring tens of parameters over a period of time (first set of proposal) or examining and decoding relatively large portions of the data stream (second one), thus introducing some delay to provide results.

With this in mind, we propose a new FR score to estimate the impact of transmission errors on the user's perceived quality: the 'piXel Loss Rate' (XLR), intuitively defined as the rate of pixels that are 'lost' (or impaired) when packets are lost during transmission. As we will show, this metric has key advantages that make it suitable as a KQI: (i) it correlates well with the (D)MOS obtained in standard subjective assessment tests; (ii) it is independent from the specific content being transmitted; and (iii) it can be estimated very robustly using a NR model that performs a mere syntactic analysis of the bitstream, thus it is simple to implement, lightweight to run, and can work with no latency (it can provide immediate predictions of the effect of the loss). Our proposal can then be seen as an evolution of PLR, focused on providing context to lost packets and enabling a short-term assessment of its real impact on quality. Hence, it is very useful for network monitoring and planning in modern IP networks. 

Finally, the work presented in this paper has two main parts: the description and justification of the FR metric, and the introduction of the NR model used to estimate the former using very lightweight network measurements, which makes it viable for its use as KQI in IP networks. Therefore, the paper is structured as follows. In Section~\ref{sec:pixel_loss_rate} we define XLR and discuss the rationale for the proposal. We also show that it is a reasonable KQI to describe the effect of network losses, as it can estimate the perceived QoE better than other alternatives. Finally, we prove that XLR is virtually independent from the transmitted content. Then, we present and validate in Section~\ref{sec:noref_est_XLR} the technique to estimate XLR based only on simple contextual information. Next, Section~\ref{sec:discussion} includes a discussion on the validity of the proposed approach, its applicability for network management, and related future work. Finally, the conclusions are presented in Section \ref{sec:conclusion}.

\begin{figure*}[h!]
\centering
\subfloat[\label{fig:dmos_vs_psnr}PSNR.]{
\includegraphics[width=0.45\textwidth]{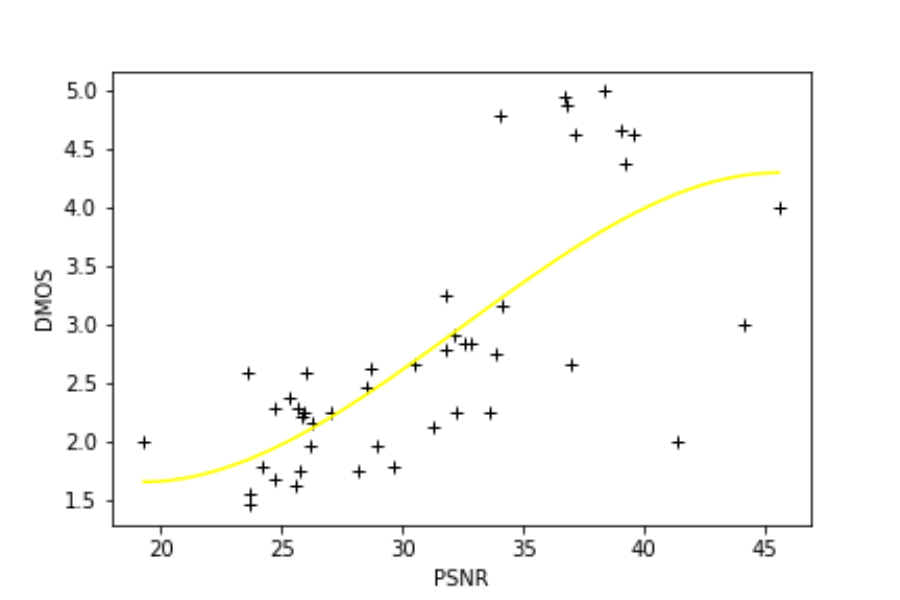}
}
\subfloat[\label{fig:dmos_vs_swissqual}Swissqual.]{
\includegraphics[width=0.45\textwidth]{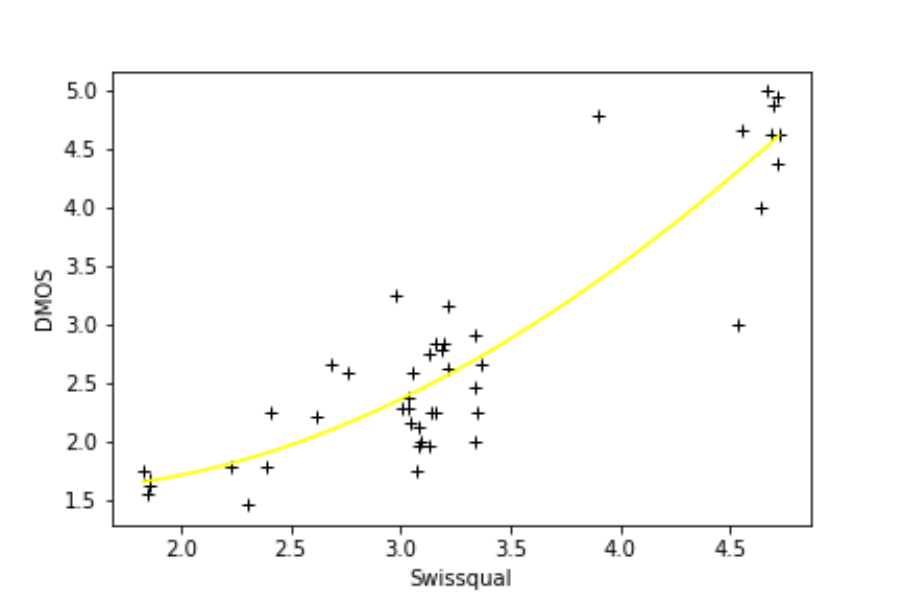}
}\\\vspace{-0.2cm}
\subfloat[\label{fig:dmos_vs_mxlr}Mean piXel Loss Rate (MXLR).]{
\includegraphics[width=0.45\textwidth]{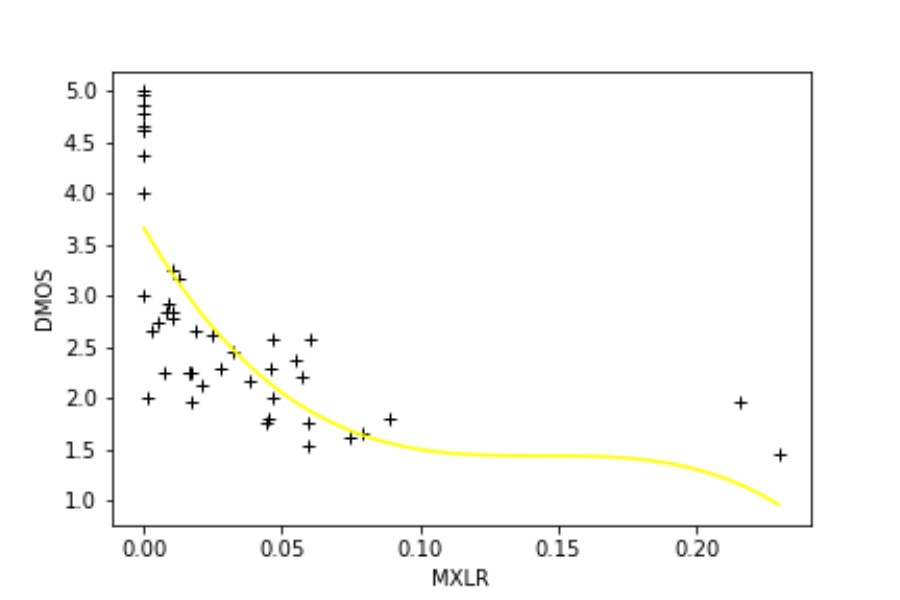}
}
\subfloat[\label{fig:dmos_vs_msxlr}Mean Squared piXel Loss Rate (MSXLR).]{
\includegraphics[width=0.45\textwidth]{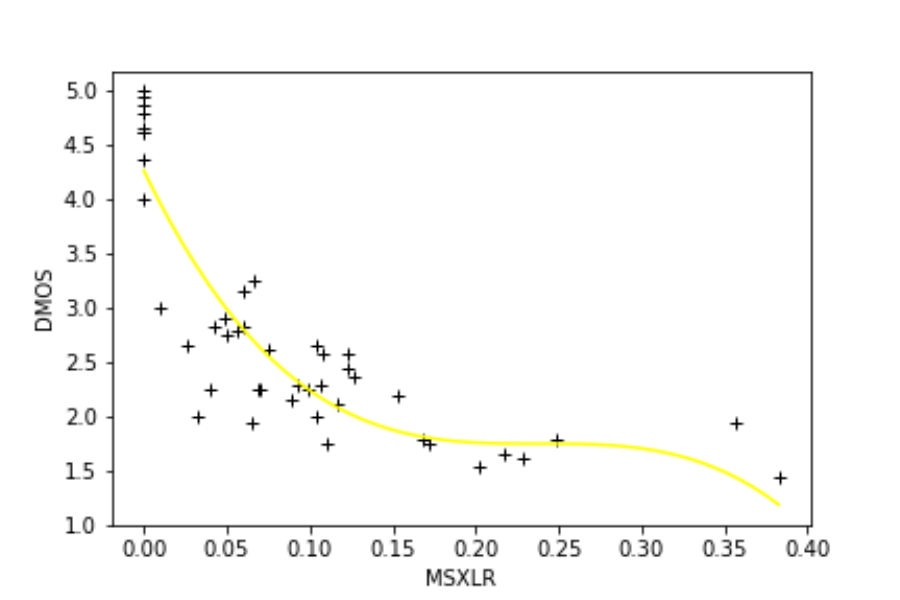}
}
\caption{\label{fig:dmos_vs_objective_metrics}Plot of the result of different objective metrics versus the subjective score in the packet-loss sequences from VQEG-HD2, where the fitting polynomial is shown in yellow.}
\vspace{-0.2cm}
\end{figure*}

\begin{table*}
\caption{\label{tab:performance_mxlr_msxlr}Performance of MXLR and MSXLR compared to PSNR and VQEG-HDTV Metrics using the VQEG-HDTV Database.}
\centering
\begin{tabular}{|c|ccccccccccc|}
\hline
\multirow{2}{*}{\textbf{Metric}} & \multirow{2}{*}{\textbf{PSNR}} & \multirow{2}{*}{\textbf{NTT}} & \multirow{2}{*}{\textbf{Opticom}} & \multirow{2}{*}{\textbf{Swissqual}} & \multirow{2}{*}{\textbf{Tektronix}} & \multirow{2}{*}{\textbf{YonseiFR}} & \textbf{YonseiRR} & \textbf{YonseiRR} & \textbf{YonseiRR} & \multirow{2}{*}{\textbf{MXLR}} & \multirow{2}{*}{\textbf{MSXLR}}\\
 &  &  &  &  &  &  & \textbf{65k} & \textbf{128k} & \textbf{256k} &  & \\
\hline 
\textbf{PCC} & 0.74 & 0.79 & 0.78 & 0.90 & 0.82 & 0.78 & 0.67 & 0.71 & 0.68 & 0.75 & 0.89\\
\hline 
\textbf{RMSE} & 0.100 & 0.100 & 0.101 & 0.071 & 0.093 & 0.102 & 0.120 & 0.114 & 0.119 & 0.107 & 0.074\\
\hline
\end{tabular}
\vspace{-.4cm}
\end{table*}

\begin{table}
\caption{\label{tab:performance_mxlr_msxlr_live_mobile}Performance of MXLR and MSXLR compared to other FR metrics using the LIVE Mobile VQA Database (Mobile).}
\centering
\begin{tabular}{|c|cccccc|}
\hline
\textbf{Metric} & \textbf{PSNR} & \textbf{SSIM} & \textbf{MSSSIM} & \textbf{VQM} & \textbf{MXLR} & \textbf{MSXLR}\\
\hline 
\textbf{PCC} & 0.70 & 0.56 & 0.68 & 0.70 & 0.68 & 0.72\\
\hline 
\textbf{RMSE} & 0.133 & 0.155 & 0.138 & 0.134 & 0.138 & 0.129\\
\hline
\end{tabular}
\vspace{-.4cm}
\end{table}

\begin{table}
\caption{\label{tab:performance_mxlr_msxlr_live_tablet}Performance of MXLR and MSXLR compared to other FR metrics using the LIVE Mobile VQA Database (Tablet).}
\centering
\begin{tabular}{|c|cccccc|}
\hline
\textbf{Metric} & \textbf{PSNR} & \textbf{SSIM} & \textbf{MSSSIM} & \textbf{VQM} & \textbf{MXLR} & \textbf{MSXLR}\\
\hline 
\textbf{PCC} & 0.76 & 0.47 & 0.80 & 0.78 & 0.73 & 0.75\\
\hline 
\textbf{RMSE} & 0.171 & 0.234 & 0.160 & 0.167 & 0.181 & 0.176\\
\hline
\end{tabular}
\vspace{-.4cm}
\end{table}

\begin{table}
\caption{\label{tab:performance_mxlr_msxlr_ughent}Performance of MXLR and MSXLR compared to other FR metrics using the AVC HD Subjective Video Database.}
\centering
\begin{tabular}{|c|cccccc|}
\hline
\textbf{Metric} & \textbf{PSNR} & \textbf{SSIM} & \textbf{MSSSIM} & \textbf{VQM} & \textbf{MXLR} & \textbf{MSXLR}\\
\hline 
\textbf{PCC} & 0.91 & 0.81 & 0.89 & 0.79 & 0.89 & 0.91\\
\hline 
\textbf{RMSE} & 0.029 & 0.040 & 0.031 & 0.041 & 0.031 & 0.028\\
\hline
\end{tabular}
\vspace{-.4cm}
\end{table}

\section{Pixel Loss Rate}
\label{sec:pixel_loss_rate}

\subsection{Definition}
Looking for the characterization of the distortion introduced by channel impairments, we propose a FR metric computed on a frame-by-frame basis, consisting in adding up the number of pixels whose resulting value differs from that of the original pixels due to transmission errors. We define this metric as piXel Loss Rate (XLR):
\begin{equation}
\text{XLR}=\frac{\sum_{i=1}^{H}\sum_{j=1}^{W}\|O(i,j)\neq D(i,j)\|}{H\cdot W}
\label{eq:xlr}
\end{equation}
where $O$ and $D$ represent the original and distorted frames (with dimensions $H$x$W$), and the term $\|E\|$ is 1 if the expression $E$ between double pipes is true and 0 otherwise.

The proposed XLR measurement does not consider the actual difference between pixel values, but only whether they remain or not equal after the transmission and decoding processes. This approach is very well suited to properly and efficiently model the effect of impairments on the decoded video sequence~\cite{2013-perez,2017-diaz_b}, as it constitutes a more general approach, far less influenced by the actual content of the sequence or, more precisely, on the motion in the scene, than existing metrics like MSE, PSNR, SSIM, or other more sophisticated objective scores, such as the ones evaluated in VQEG-HDTV project~\cite{2010_vqeg}. Furthermore, the XLR metric does not depend on the error concealment mechanism or on the actions of the decoder when facing desynchronization. The reason is that, whichever the technique used, the new pixel values will be virtually always different from the original ones, except in an extremely low number of occurrences in a handful of cases (e.g. cartoons or animated movies with little motion). Therefore, the number of pixels that are different in both sequences will in practice be the same regardless of the error concealment mechanism.

\subsection{Generalization: temporal pooling}
XLR is a metric on a single frame. When applied to a frame structure, such as a prediction structure, a sequence or a segment of video, its mean value MXLR will be used as global measurement of the distortion:
\begin{equation}
\text{MXLR}=\frac{1}{N_{\text{F}}}\sum_{i=1}^{N_{\text{F}}}\text{XLR}_{f_{i}}
\end{equation}
where $f_{a}$, $f_{b}$, ..., $f_{N_{\text{F}}}$ are the $N_{\text{F}}$ frames that make up this structure. Additionally, we will also evaluate the mean value of the square root of XLR (MSXLR), as, for other video quality effects, it is known that the quality is related to the linear dimension of the frame~\cite{2011-cermak}. 
\begin{equation}
\text{MSXLR}=\frac{1}{N_{\text{F}}}\sum_{i=1}^{N_{\text{F}}}\sqrt{\text{XLR}_{f_{i}}}
\end{equation}

\subsection{Pixel Loss Rate as Key Quality Indicator}
\label{sec:xlr_as_kqi}

\subsubsection{Experimental Design}
To understand whether XLR is a reasonable KQI for video packet losses, we have evaluated its performance as FR metric for video quality, comparing it with other FR pixel-based metrics using three well-known widely-used databases: (i) the VQEG-HDTV Database, created, validated and used by many labs worldwide during the HDTV project~\cite{2010_vqeg} (specifically, we have used the subset VQEG-HD2, made publicly available by IRCyNN~\cite{2010_barkowsky}); (ii) the LIVE Mobile VQA Database generated by the University of Texas at Austin~\cite{2012-moorthy}; and (iii) the AVC HD Subjective Video (AHSV) Database developed by the University of Ghent~\cite{2013-staelens}. They are briefly described next.

The Video Quality Experts Group (VQEG) performed a series of experiments under its HTDV project, with the aim of standardizing objective metrics that could characterize transmission errors, among other artifacts. We have tested XLR against the same sequences, and compared the results with the ones reported by VQEG. The report includes on the one hand the objective scores from a number of metrics (PSNR and eight test metrics that can be estimated from NR approaches) that were evaluated in the project, which can be considered state-of-the-art for the prediction of video quality under coding and transmission impairments. On the other hand, VQEG also provides the MOS's of all the sequences. We have considered all the Hypothetical Reference Circuits (HRCs) in the dataset related to network impairments: \begin{itemize}
\item HRC1: No packet loss. Used as reference. 
\item HRC5: Short burst, 0.7\%~PLR. 
\item HRC6: Long burst, 4.2\%~PLR. 
\item HRC7: Short burst, 0.7\%~PLR. 
\item HRC8: Short burst, 0.7\%~PLR. 
\end{itemize}
There are 15 different Source sequences (SRCs) for each of those HRCs, giving a total of 75 Processed Video Sequences (PVS's) under evaluation. All of them are 1080i, H.264/AVC, QP~26, 13.5~Mbps. Those sequences have been used to compute the XLR, comparing the values of HRC5-HRC8 to the values of HRC1, which is considered the original sequence. However, the sequences available in the repository are not in pure raw format, but encoded with a perceptually-lossless AVC compression. Therefore, we estimate the XLR value of a single frame as: 
\begin{equation}
\overline{\text{XLR}}=\frac{\sum_{i=1}^{H}\sum_{j=1}^{W}\|\left|O(i,j)-D(i,j)\right|<Q\|}{H\cdot W}
\label{eq:xlr_encoded_ref}
\end{equation}
where we have used $Q=16$, which means that we are considering differences in the 4 highest significant bits of an 8-bit difference image.

The LIVE Mobile VQA Database includes 10~1280x720-resolution SRCs and 4~HRCs corresponding with the lossy transmission of different encoded versions of the sources through as many wireless channels. Thus, there are 40~PVS's under evaluation. In the same way as before, the scores of the encoded versions were used as reference and the XLR of every frame was computed using~\eqref{eq:xlr_encoded_ref}. Additionally, the database includes the subjective scores of all the sequences. In this case, however, they employed two types of displays during the subjective assessment: mobiles and tablets. Thus, they provide the subjective scores separately. As they do not include results for any specific objective metrics, we have compared the performance of XLR to that of widely-used general-purpose objective FR metrics: PSNR, SSIM, MSSSIM, and VQM.

Finally, the AVC HD Subjective Video Database contains 9~H.264/AVC-encoded versions of 8~different sources of a resolution of 1920x1080 pixels. Additionally, network impairments are simulated by dropping single and multiple slices in the encoded video files, resulting in 48~HRCs. Therefore, the dataset contains 384~PVS's. Again, the unimpaired versions are used as reference in our experiments. In the same way as in the other databases, subjective scores for all PVS's are provided. As before, since results for specific objective metrics are not included, we have compared the performance of XLR to that of PSNR, SSIM, MSSSIM, and VQM.

\begin{figure}[t!]
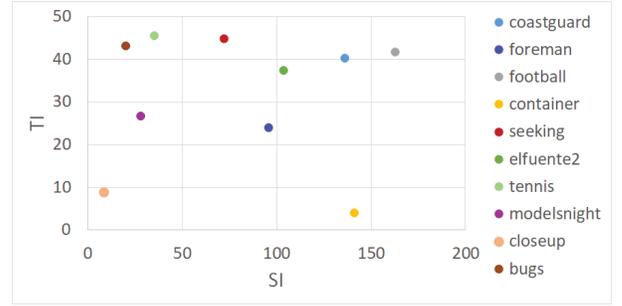

\centering
\includegraphics[width=0.9\columnwidth]{{{Figures/siti}}}
\caption{\label{fig:siti}SI and TI of the sequences used to validate the independence of XLR from video content.}
\vspace{-0.5cm}
\end{figure}

\begin{figure*}[th!]
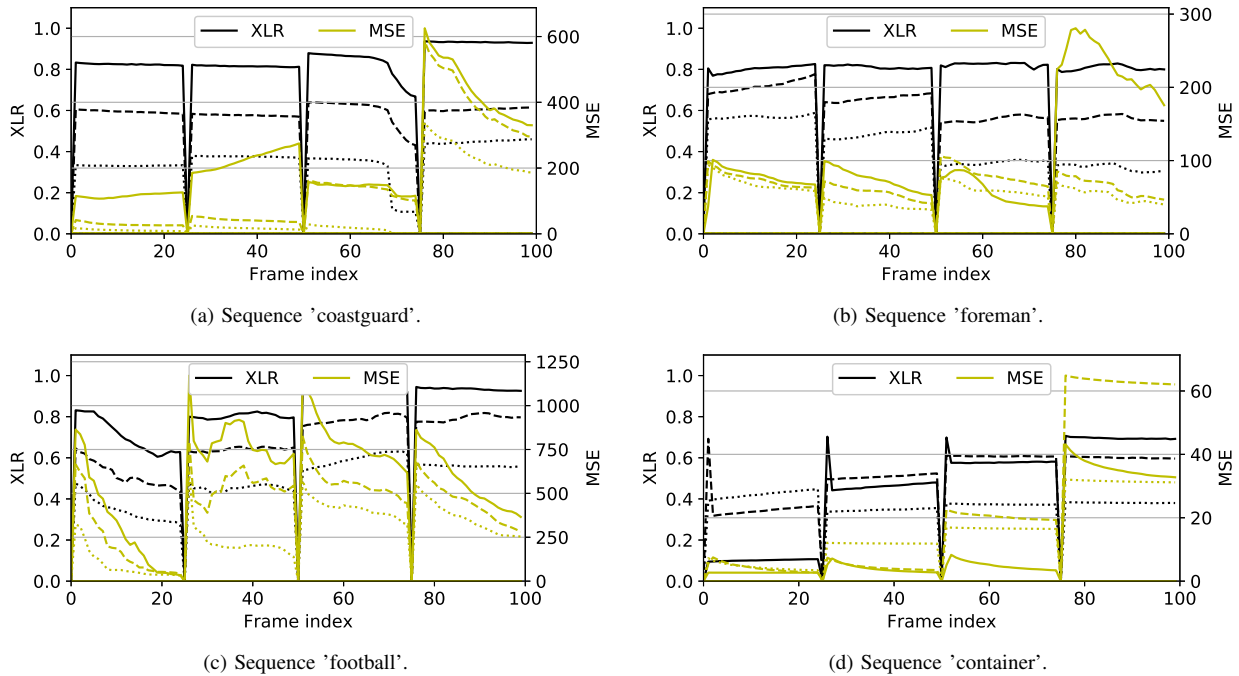

\centering
\subfloat[\label{fig:study_error_propagation_affected_pixels_coastguard_iperiod_25}Sequence 'coastguard'.]{
\includegraphics[width=0.45\textwidth]{{{Figures/pixel_loss_rate_coastguard_352x288_25fps_br_2000000_Iperiod_25_B_0_ref_1_JM.ts_packet_drop_analysis}}}
}
\subfloat[\label{fig:study_error_propagation_affected_pixels_foreman_iperiod_25}Sequence
'foreman'.]{
\includegraphics[width=0.45\textwidth]{{{Figures/pixel_loss_rate_foreman_352x288_25fps_br_2000000_Iperiod_25_B_0_ref_1_JM.ts_packet_drop_analysis}}}
}\\
\vspace{-2mm}
\subfloat[\label{fig:study_error_propagation_affected_pixels_football_iperiod_25}Sequence
'football'.]{
\includegraphics[width=0.45\textwidth]{{{Figures/pixel_loss_rate_football_352x240_25fps_br_2000000_Iperiod_25_B_0_ref_1_JM.ts_packet_drop_analysis}}}
}
\subfloat[\label{fig:study_error_propagation_affected_pixels_container_iperiod_25}Sequence
'container'.]{
\includegraphics[width=0.45\textwidth]{{{Figures/pixel_loss_rate_container_352x288_25fps_br_2000000_Iperiod_25_B_0_ref_1_JM.ts_packet_drop_analysis}}}
}
\caption{\label{fig:study_error_propagation_affected_pixels_iperiod_25}Propagation of the error caused by single packet losses throughout the sequence for different QCIF sequences encoded with an IDR period of 25 frames at 2 Mbps. There are 3 pairs of curves per sequence, each of the pairs representing the propagation of the error caused by dropping a different packet belonging to the first P-frame in the structure measured in terms of XLR (black lines) and MSE (yellow lines). Furthermore, each pair is represented using a different line style, solid, dashed and dotted, respectively according to how close they are to the beginning of the frame.}
\vspace{-5mm}
\end{figure*}

\subsubsection{Model Evaluation}
The performance of the different metrics has been evaluated by using the same procedure described by VQEG in its HDTV project~\cite{2010_vqeg}. First, the Differential Mean Opinion Score (DMOS) is used as ground truth for the quality of each of the processed video sequences. Secondly, the mentioned report states: 'Subjective rating data often are compressed at the ends of the rating scales. It is not reasonable for objective models of video quality to mimic this weakness of subjective data. Therefore, a non-linear mapping step was applied before computing any of the performance metrics, by fitting the metrics to a cubic polynomial, which has been found to perform well empirically.' So, we have implemented such mapping using Sequential Least SQuares Programming (SLSQP)~\cite{1988-kraft}.

First, as visual examples, Fig.~\ref{fig:dmos_vs_objective_metrics} shows the relationship between the objective score and the DMOS, together with the non-linear mapping curve, for four different metrics: PSNR and Swissqual (as example of a VQEG-HDTV proposed metric), MXLR, and MSXLR. It can be seen that both XLR-based metrics capture the general behavior of the subjective scores with different levels of impairment.

Furthermore, after the above mentioned polynomial mapping, the Pearson Correlation Coefficient (PCC) and the Root Mean Square Error (RMSE) have been computed and the results are shown in Tables~\ref{tab:performance_mxlr_msxlr},~\ref{tab:performance_mxlr_msxlr_live_mobile},~\ref{tab:performance_mxlr_msxlr_live_tablet}, and~\ref{tab:performance_mxlr_msxlr_ughent}, respectively for the different sequences. It can be observed that the MXLR performance is already similar to that of PSNR and other metrics. Moreover, its linear-dimension value (MSXLR) correlates well with the users' rates and its performance is comparable to that of more complex state-of-the-art objective metrics estimable or not via bitstream analysis. These results mean that, even though XLR is not able to capture all the factors that impact the video QoE and so predict it completely accurately, despite its simplicity, its performance is comparable to that of other widely-used far more complex FR metrics for estimating the perceived quality of videos. This proves that it is able to capture the component in videos degraded by packet losses that impacts the most on the user perception.

The differences between datasets are mainly due not only to the number of PVS's included, but also to the nature and impact of the impairments introduced. In this context, for instance, most errors introduced in the AHSV Database have a moderate impact on the quality of the resulting sequence, considering it as a whole. Therefore, the range of scores of all the metrics is relatively small and hence their differences are also less significant. In this respect, the VQEG-HDTV database constitutes the more complete and varied one.

Finally, it is important to note that, as the results suggest, even though there exist a strong relationship between XLR and QoE, this relation is not linear. Hence, we have tentatively proposed MSXLR as aggregated XLR-based metric, achieving significantly better results. However, other pooling functions could be used.

\begin{figure*}[h!]
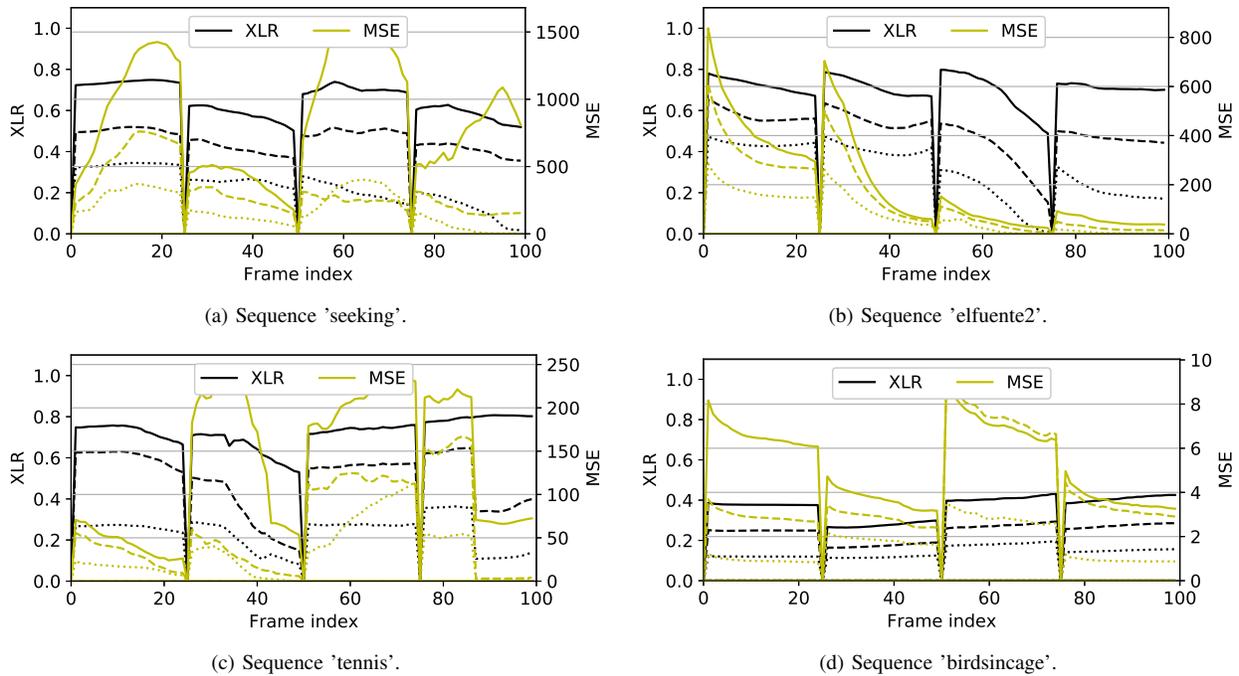

\centering
\subfloat[\label{fig:study_error_propagation_affected_pixels_seeking_iperiod_25_fhd}Sequence 'seeking'.]{
\includegraphics[width=0.45\textwidth]{{{Figures/pixel_loss_rate_seeking_1920x1080_25fps_br_8000000_Iperiod_25_sc_B_0_ref_1.ts_packet_drop_analysis}}}
}
\subfloat[\label{fig:study_error_propagation_affected_pixels_elfuente2_iperiod_25_fhd}Sequence
'elfuente2'.]{
\includegraphics[width=0.45\textwidth]{{{Figures/pixel_loss_rate_elfuente2_1920x1080_30fps_br_8000000_Iperiod_25_sc_B_0_ref_1.ts_packet_drop_analysis}}}
}\\
\vspace{-2mm}
\subfloat[\label{fig:study_error_propagation_affected_pixels_tennis_iperiod_25_fhd}Sequence
'tennis'.]{
\includegraphics[width=0.45\textwidth]{{{Figures/pixel_loss_rate_tennis_1920x1080_24fps_br_8000000_Iperiod_25_sc_B_0_ref_1.ts_packet_drop_analysis}}}
}
\subfloat[\label{fig:study_error_propagation_affected_pixels_birdsincage_iperiod_25_fhd}Sequence
'birdsincage'.]{
\includegraphics[width=0.45\textwidth]{{{Figures/pixel_loss_rate_birdsincage_1920x1080_30fps_br_8000000_Iperiod_25_sc_B_0_ref_1.ts_packet_drop_analysis}}}
}
\caption{\label{fig:study_error_propagation_affected_pixels_iperiod_25_fhd}Propagation of the error caused by single packet losses throughout the sequence for different Full HD sequences encoded with an IDR period of 25 frames at 8 Mbps. There are 3 pairs of curves per sequence, each of the pairs representing the propagation of the error caused by dropping a different packet belonging to the first P-frame in the structure measured in terms of XLR (black lines) and MSE (yellow lines). Furthermore, each pair is represented using a different line style, solid, dashed and dotted, respectively according to how close they are to the beginning of the frame.}
\vspace{-5mm}
\end{figure*}

\begin{figure*}[h!]
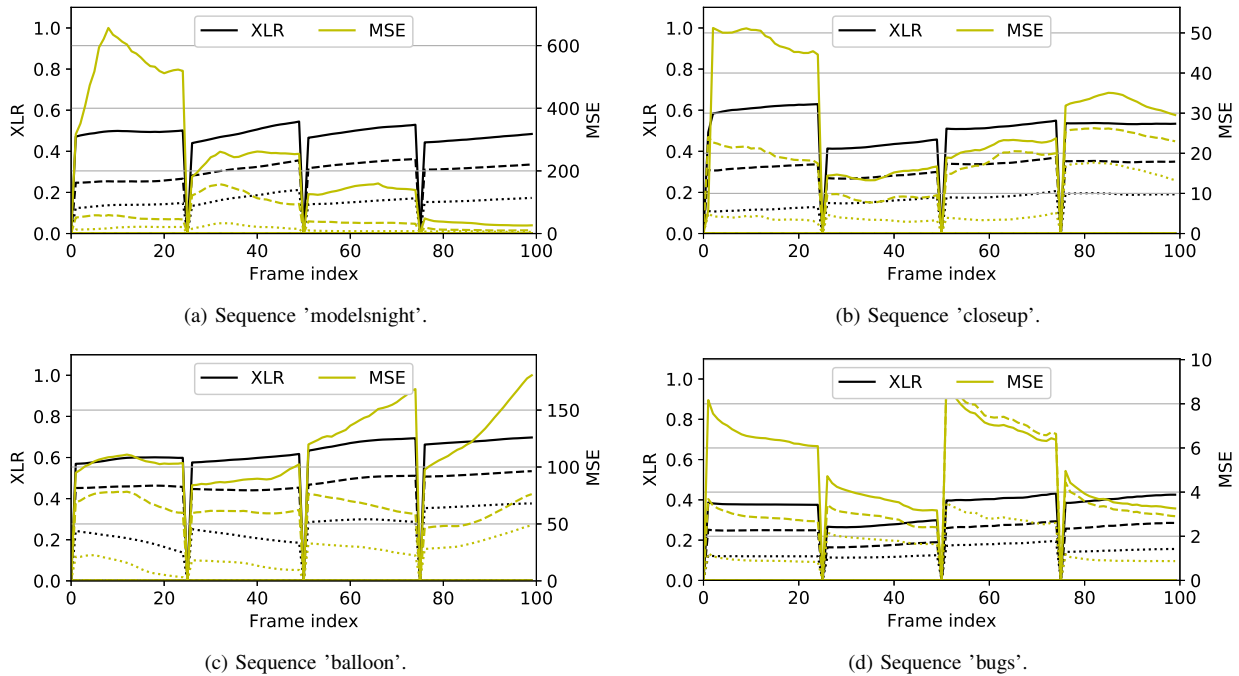

\centering
\subfloat[\label{fig:study_error_propagation_affected_pixels_seeking_iperiod_25_4k}Sequence 'modelsnight'.]{
\includegraphics[width=0.45\textwidth]{{{Figures/pixel_loss_rate_modelsnight_4096x2160_50fps_br_16000000_Iperiod_25_sc_B_0_ref_1.ts_packet_drop_analysis}}}
}
\subfloat[\label{fig:study_error_propagation_affected_pixels_elfuente2_iperiod_25_4k}Sequence
'closeup'.]{
\includegraphics[width=0.45\textwidth]{{{Figures/pixel_loss_rate_closeup_4096x2160_50fps_br_16000000_Iperiod_25_sc_B_0_ref_1.ts_packet_drop_analysis}}}
}\\
\vspace{-2mm}
\subfloat[\label{fig:study_error_propagation_affected_pixels_tennis_iperiod_25_4k}Sequence
'balloon'.]{
\includegraphics[width=0.45\textwidth]{{{Figures/pixel_loss_rate_balloon_4096x2160_50fps_br_16000000_Iperiod_25_sc_B_0_ref_1.ts_packet_drop_analysis}}}
}
\subfloat[\label{fig:study_error_propagation_affected_pixels_birdsincage_iperiod_25_4k}Sequence
'bugs'.]{
\includegraphics[width=0.45\textwidth]{{{Figures/pixel_loss_rate_birdsincage_1920x1080_30fps_br_8000000_Iperiod_25_sc_B_0_ref_1.ts_packet_drop_analysis}}}
}
\caption{\label{fig:study_error_propagation_affected_pixels_iperiod_25_4k}Propagation of the error caused by single packet losses throughout the sequence for different 4K sequences encoded with an IDR period of 25 frames at 16 Mbps. There are 3 pairs of curves per sequence, each of the pairs representing the propagation of the error caused by dropping a different packet belonging to the first P-frame in the structure measured in terms of XLR (black lines) and MSE (yellow lines). Furthermore, each pair is represented using a different line style, solid, dashed and dotted, respectively according to how close they are to the beginning of the frame.}
\vspace{-5mm}
\end{figure*}

\subsection{Independence from video content}
As above-mentioned, the results of XLR virtually do not depend on the content of the sequence. To prove it, we use three sets of four 100-frame sequences. Each set contains clips of the same resolution (QCIF, Full HD or 4K) presenting very different characteristics from the point of view of content and spatial and temporal complexity, as shown in Figure~\ref{fig:siti}, which depicts the Spatial Information (SI) and Temporal Information (TI) indicators of the 12 sequences. All the clips were extracted from publicly available video datasets, respectively~\cite{2002-cipr,2016-netflix,2019-blackmagic}. The three sets of videos where encoded with the reference software H.264/AVC encoder (v. JM 19.0) at respectively 2, 8 and 16~Mbps using an Instantaneous Decoder Refresh (IDR) period of 25~frames following the pattern IPP... where intra-MBs are introduced in predictive frames whenever they are more efficient than inter-MBs working in default mode, so as to enable a more real setup. Moreover, only one slice is used per frame. Therefore, all twelve encoded sequences were made up of four consecutive independent prediction structures. Figures~\ref{fig:study_error_propagation_affected_pixels_iperiod_25},~\ref{fig:study_error_propagation_affected_pixels_iperiod_25_fhd}, and~\ref{fig:study_error_propagation_affected_pixels_iperiod_25_4k} depict the evolution of the error propagation caused by the loss of just one packet at a time. Every graph shows the four consecutive structures of frames of one sequence. Every structure includes 3 pairs of curves, each of the pairs representing the propagation of the error caused by dropping a different packet belonging to the first P-frame in the structure measured in terms of XLR (black lines) and MSE (yellow lines). Furthermore, each pair is represented using a different line style, solid, dashed and dotted, respectively according to how close they are to the beginning of the frame. So, in total we have analyzed the impact of losing 3 different packets per structure, resulting in 12 pairs of lines per graph. MSE was chosen as it is the most commonly used in similar studies in the literature~\cite{2006-he, 2009-li, 2010-zhang}. However, the conclusions can be extended to other metrics.
%the superimposition of the results of singly dropping 12 different packets, three per prediction structure, all of them transporting information of the first P-frame in the structure. 
%The distortion is shown in terms of XLR (blue solid lines) and MSE (red dashed lines), as this metric is the most commonly used in similar studies in the literature~\cite{2006-he, 2009-li, 2010-zhang}. Although we have used MSE in the experiments, the conclusions can be extended to other metrics.

Observing the figures, one can quickly see the differences between the two scores. Regarding XLR, the percentage of pixels per frame whose value differs from the original one remains fairly constant with the distance to the frame where the error takes place, or goes up or down very slowly over time. These tendencies remain steady throughout the prediction structure except for the few cases where a major disturbance in the scene (e.g. considerable increment of motion in the vertical axis, scene change...) gets in the way. On the contrary, the results in terms of MSE show a more unpredictable, changing behavior, much more influenced by the transmitted content. Furthermore, this behavior varies notably from structure to structure. Often, but definitively not in most cases, it follows the fading behavior frequently mentioned in the literature~\cite{2006-he, 2009-li, 2010-zhang}. However, it must be noted that the conditions of the experiments referred in the papers differ from ours mainly in terms of the
%fact that the characteristics of an important element differ. Indeed, besides the motion in the content and the metric used to measure the error, the way the distortion evolves with the distance between the analyzed frame and the one where the error takes place highly relies on the
ratio between intra- and inter-mode basic processing units (e.g. macroblocks -MBs- in H.264/AVC and coding tree units -CTUs- in H.265/HEVC) in the predicted pictures. Certainly, the greater the proportion of intra-mode MBs/CTUs used to encode the pictures in the sequence, the quicker the error fades away, and therefore the steeper the curve is. This number typically depends on the motion of the content, if encoders are used in default mode. The more motion in the scene, the greater the difference between reference and predicted frames and therefore the greater the number of intra-mode units required. However, the proportion can be varied if the sensibility of the selector between modes is modified or if intra-refresh or a similar mode is employed to artificially increase the number of intra-mode units. In this respect, in many existing studies, the number of intra-mode MBs/CTUs is significantly greater than the one typically used in everyday video streaming services, much closer to the one that results from running the encoders in default mode, as carried out in our experiments.

In summary, XLR presents a practically independent behavior from the content, particularly if all dependent frames belong to the same take, which is overwhelmingly the case considering the relatively small size of typical prediction structures compared to the length of typical takes. Under this premise, with high probability, no big irregularities will break the steady tendency. Of course, the probability is even higher if the scene change detection mode is enabled when encoding the sequence.

\section{No-Reference Estimation of XLR}
\label{sec:noref_est_XLR}
 
XLR can be estimated directly from the bitstream without using a reference sequence. That is, the FR metric can be extended to a more versatile and useful NR method. The rationale behind it is that the fraction of pixels that are different between the original and the distorted versions of the transmitted sequence is equivalent to the percentage of frame area that is affected by packet losses. The latter value can be estimated considering the unequal impact of the lost packets in the bitstream depending of their location along this flow and the propagation of errors within interdependent frames in the video sequence. 

Therefore, to derive a proper NR technique to estimate the XLR in the received packet stream, we first characterize the distortion caused by packets losses, as mentioned, in terms of the area of those frames that is impaired. This characterization is based on a preliminary analysis and formulation carried out previously targeting the formalization of a stochastic packet-level distortion model~\cite{2017-diaz_b}.

Initially, the case where only one network packet is lost is discussed and formalized. Later on, this result is generalized for an arbitrary number of packet losses. 

\begin{figure}[t]
\centering
\subfloat[\label{fig:error_propagation_impact_IPP_one_packetloss}Prediction structure IPP... in display order.]{
\includegraphics[width=0.47\textwidth]{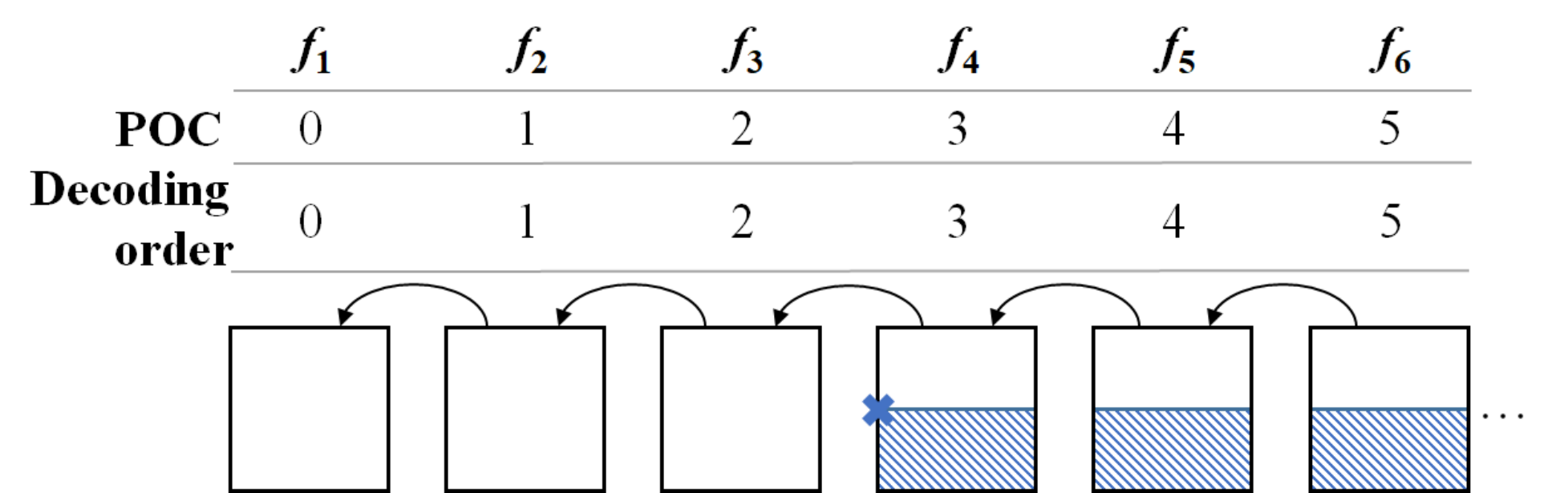}
}\\
\subfloat[\label{fig:error_propagation_impact_hier_one_packetloss}Prediction structure $\text{IB}_{2}\text{B}_{1}\text{B}_{2}\text{P}$... in display order.]{
\includegraphics[width=0.47\textwidth]{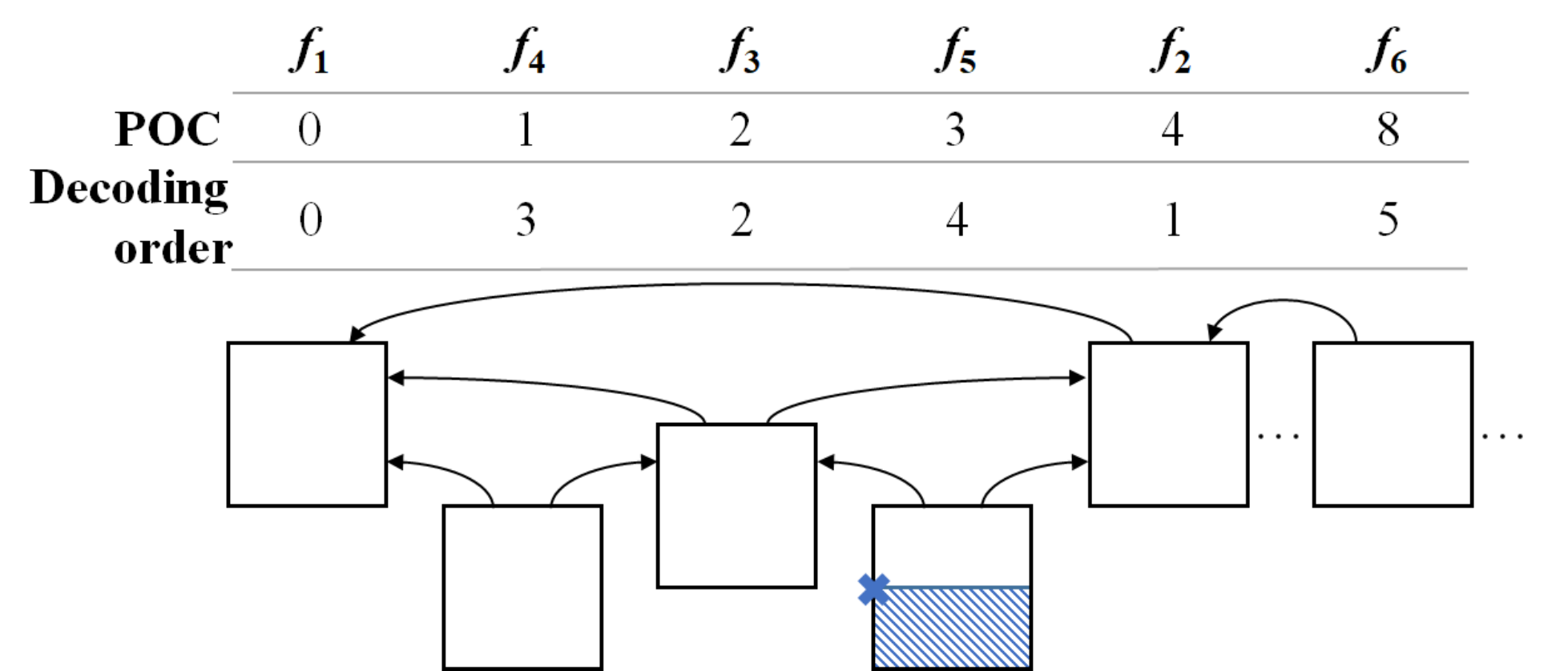}
}
\caption{\label{fig:error_propagation_single_loss}Graphic representation of the distortion introduced as a result of the propagation of the error caused by the loss of a single packet. Figures include the picture order count (POC) values and
decoding order for each picture. Arrows point at the direct reference of each frame. The blue cross indicates the initial pixel in the frame about which the lost packet carries information.}
\vspace{-0.5cm}
\end{figure}

\subsection{Impact of a Single Packet Loss}

\subsubsection{Discussion}
The model directly derives from the analysis of the propagation of the error caused by the loss of a packet in a frame to the frames that depend on it, included in Section~\ref{sec:pixel_loss_rate}. It will be assumed that all interdependent pictures belong to the same scene, where errors can be presumed to propagate affecting a rather constant number of pixels in all the dependent frames.

\subsubsection{Formalization}

The impaired area in dependent frames is modeled assuming that it remains constant with the distance. Thus, the estimated XLR value introduced by the loss of packet $p_{f}$, which carries information belonging to frame $f$, is modeled as follows:
\begin{equation}
\widehat{\text{XLR}}_{f,p_{f}}=\xi_{f,p_{f}}
\end{equation}
where $\xi_{f,p_{f}}$ is the fraction of impaired pixels in frame $f$ when packet $p_{f}$ is the only one lost in the window of observation.

Similarly, for all the frames in this window that directly or indirectly depend on frame $f$ for decoding, $f_i$, we estimate their XLR value as:
\begin{equation}
\widehat{\text{XLR}}_{f_i,p_{f}} \approx \widehat{\text{XLR}}_{f,p_{f}}=\xi_{f,p_{f}}
\end{equation}

Fig.~\ref{fig:error_propagation_single_loss} presents two schematic examples of the distortion introduced as a result of the propagation of the error caused by the loss of a single packet. In the first one (Fig.~\ref{fig:error_propagation_impact_IPP_one_packetloss}), we assume that all inter frames have only one direct reference: the previous frame. Also, to show the effect of multiple direct references, Fig.~\ref{fig:error_propagation_impact_hier_one_packetloss} includes an example where a hierarchical temporal prediction structure is used to encode the content. The frames are shown in display order. In both figures, it is shown that the isolated loss of packet $p_{f}$ in frame $f$ results in $\xi_{f,p_{f}}$ percent of the pixels not only in this frame but also in those ones depending on it taking wrong values. Thus, $\xi_{f,p_{f}}$ denotes how potentially distorting packet $p_{f}$ is.

Assuming that the spatial complexity of each frame in the sequence is rather uniform and that the whole image is encoded using equivalent encoding parameter values (particularly regarding the quantization parameter -QP-), we can consider that the size of the packets of a given frame is proportional to the number of pixels whose information is transported by it. Therefore, the index of the lost packet, jointly with the size of the packets belonging to frame $f$, determines the percentage of the impaired picture. If $s_{f,1}$, $s_{f,2}$, ..., $s_{f,N_{\text{P},f}}$ is the size in number of octets of the $N_{\text{P},f}$ packets carrying information of frame $f$, as they are located along the transmission stream, the impaired area $\xi_{f,p_{f}}$ can be estimated as:
\begin{equation}
\xi_{f,p_{f}}=\frac{\sum_{i=p_{f}}^{N_{p,f}}s_{f,i}}{\sum_{i=1}^{N_{p,f}}s_{f,i}}
\label{eq:impaired_area_frame}
\end{equation}

This expression arises from the loss of synchronization taking place whenever a packet is loss. In this situation, the decoder discards the bitstream after the loss until the following synchronization marker arrives, which happens by the arrival of the next Network Abstraction Layer (NAL) unit header, i.e., by the beginning of the subsequent frame/slice. Therefore, all the data between the loss and the new marker is considered to also be lost for all purposes.

\begin{figure}[t]
\centering
\subfloat[\label{fig:error_propagation_impact_IPP_two_packetlosses_a}IPP... in display order; $\xi_{f_{a},p_{f_{a}}}<\xi_{f_{b},p_{f_{b}}}$.]{
\includegraphics[width=0.47\textwidth]{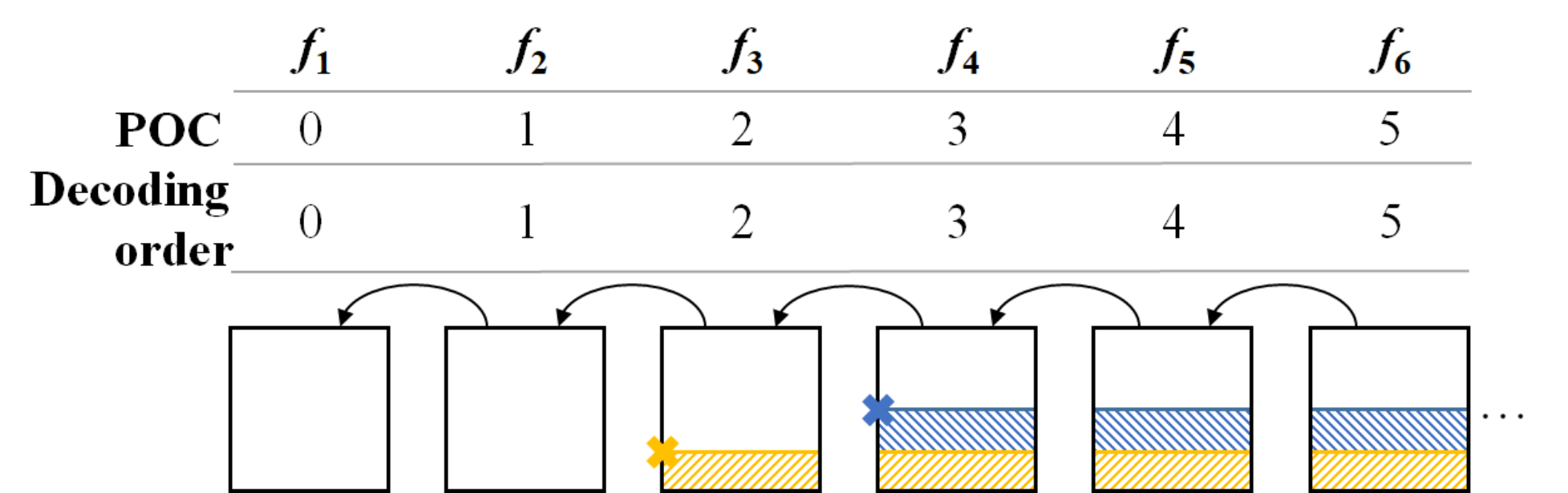}
}\\
\subfloat[\label{fig:error_propagation_impact_IPP_two_packetlosses_b}IPP... in display order; $\xi_{f_{a},p_{f_{a}}}\protect\geq\xi_{f_{b},p_{f_{b}}}$.]{
\includegraphics[width=0.47\textwidth]{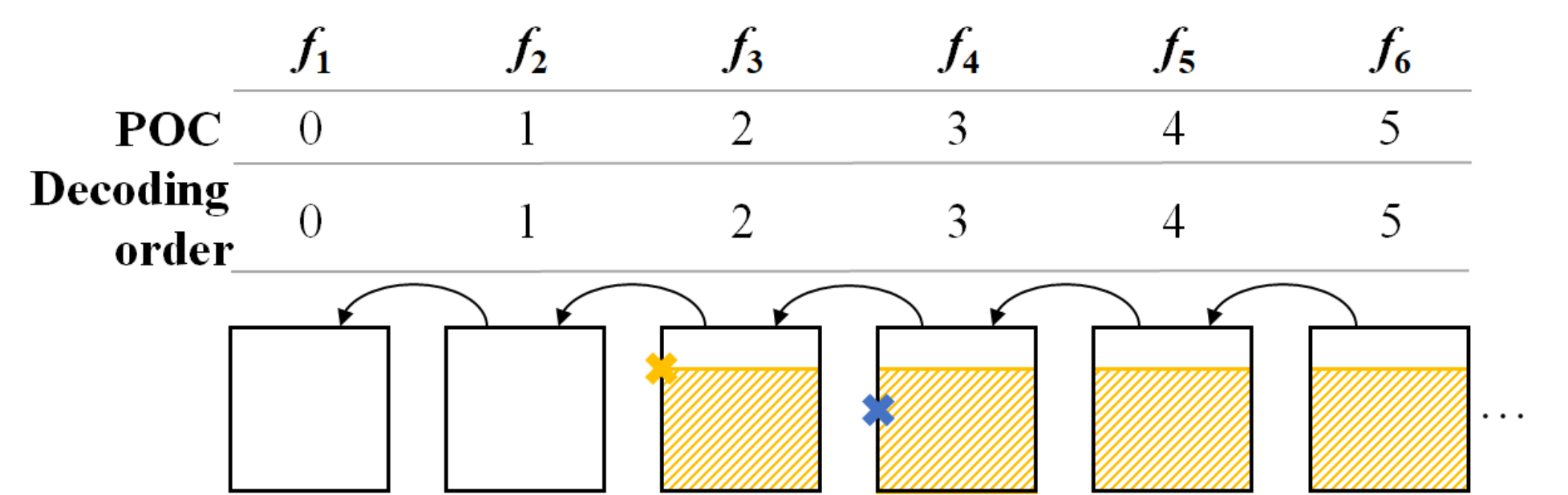}
}\\
\subfloat[\label{fig:error_propagation_impact_hier_two_packetlosses_a}$\text{IB}_{2}\text{B}_{1}\text{B}_{2}\text{P}$... in display order; $\xi_{f_{a},p_{f_{a}}}<\xi_{f_{b},p_{f_{b}}}$.]{
\includegraphics[width=0.47\textwidth]{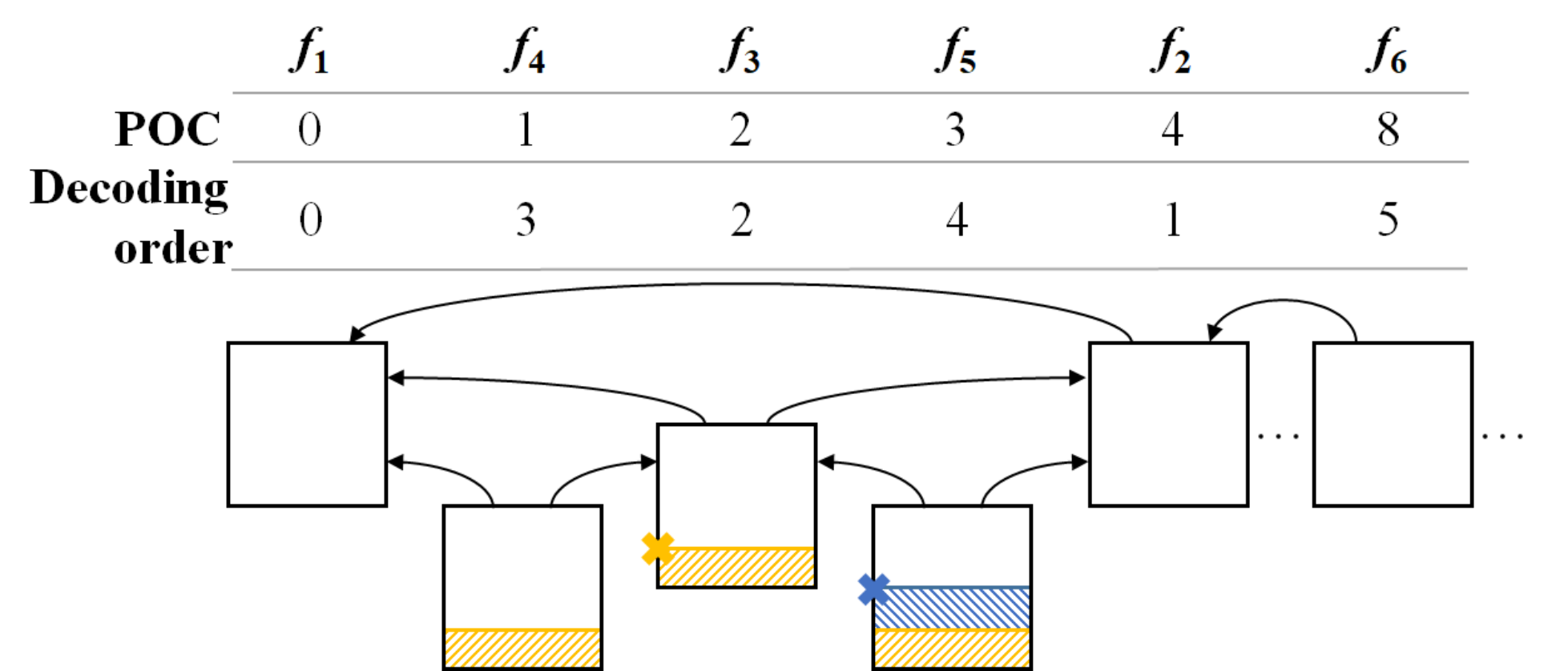}
}\\
\subfloat[\label{fig:error_propagation_impact_hier_two_packetlosses_b}$\text{IB}_{2}\text{B}_{1}\text{B}_{2}\text{P}$... in display order; $\xi_{f_{a},p_{f_{a}}}\protect\geq\xi_{f_{b},p_{f_{b}}}$.]{
\includegraphics[width=0.47\textwidth]{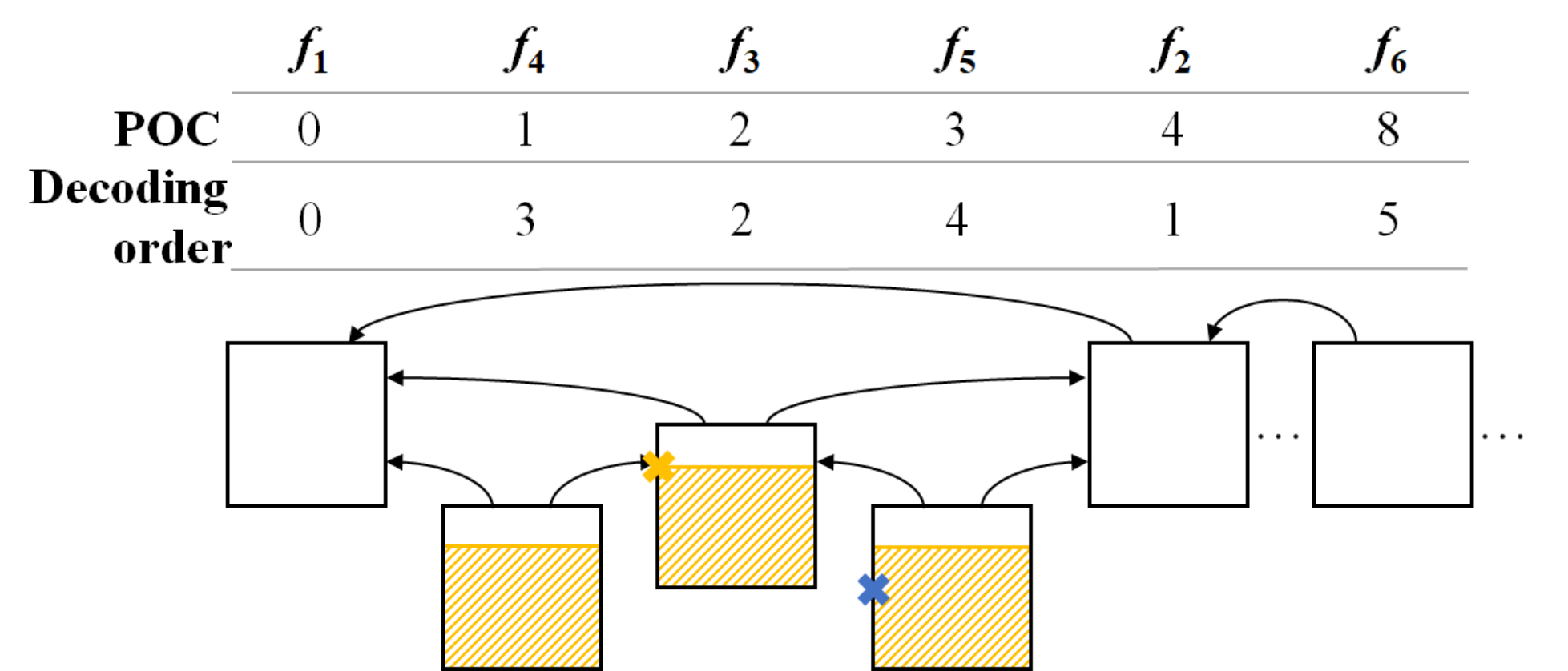}
}

\caption{\label{fig:error_propagation_impact_two_packetlosses}Graphic representation of
the distortion introduced as a result of the propagation of the error
caused by the loss of two packets. Figures include the POC values and
decoding order for each picture. Arrows point at the direct reference
of each frame. The orange cross indicates the initial pixel in the
frame the first lost packet carries information about and the blue
one indicates the initial pixel in the frame the second lost packet
carries information about.}
\vspace{-0.7cm}
\end{figure}

\begin{figure}[t]
\centering
\subfloat[\label{fig:error_propagation_impact_IPP_three_packetlosses}Prediction structure IPP... in display order.]{
\includegraphics[width=0.47\textwidth]{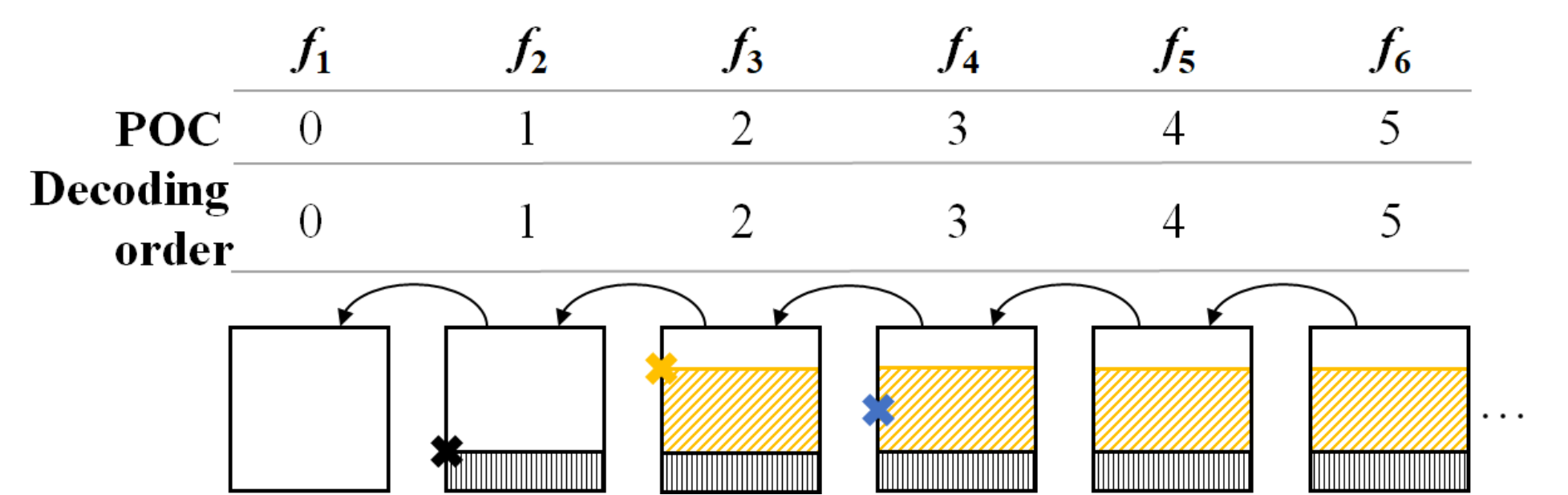}
}\\
\subfloat[\label{fig:error_propagation_impact_hier_three_packetlosses}Prediction structure $\text{IB}_{2}\text{B}_{1}\text{B}_{2}\text{P}$... in display order.]{
\includegraphics[width=0.47\textwidth]{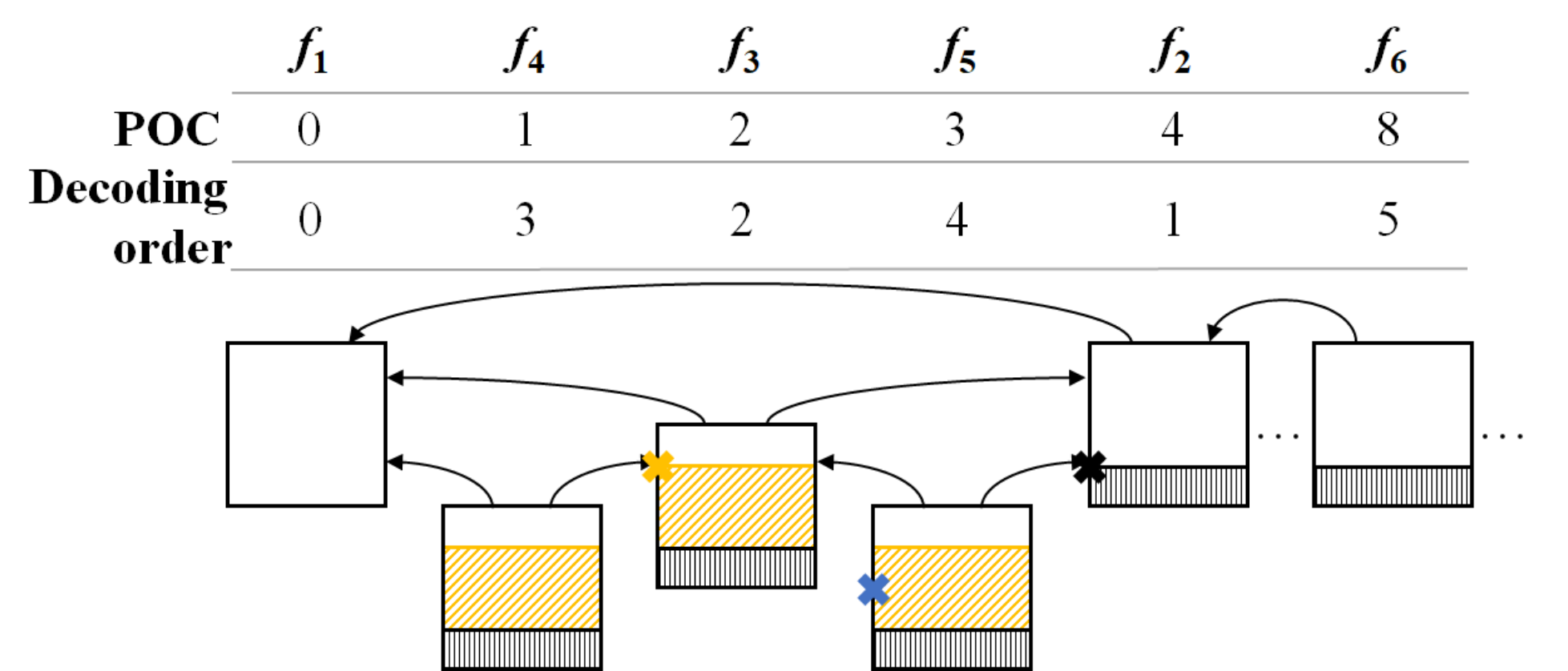}
}
\caption{\label{fig:error_propagation_impact_three_packetlosses}Graphic representation
of the distortion introduced as a result of the propagation of the
error caused by the loss of three packets. Figures include the POC values and
decoding order for each picture. Arrows point at the direct
reference of each frame. The black cross indicates the initial pixel
in the frame the first lost packet carries information about, the
orange cross indicates the initial pixel in the frame the second lost
packet carries information about and the blue one indicates the initial pixel in the frame the third lost packet carries information about.}
\vspace{-0.5cm}
\end{figure}

\subsection{Impact of Several Simultaneous Packet Losses}

\subsubsection{Discussion}

We now analyze the effect of simultaneously losing two or more packets belonging to the same or different interrelated frames.

We first consider the loss of two packets. To that end, let packet $p_{f_{a}}$ belonging to frame $f_{a}$ be the first of the two lost packets in the transmitted packet stream and packet $p_{f_{b}}$ in frame $f_{b}$ be the second one. According to the previous subsection, the isolated loss of the first one would affect $\xi_{f_{a},p_{f_{a}}}$ percent of the pixels of frame $f_{a}$ and of the frames depending on it, whereas the isolated loss of the second one would change the value of $\xi_{f_{b},p_{f_{b}}}$ percent of the pixels of frame $f_{b}$ and of the frames relying on it. Two different cases, illustrated
in Fig.~\ref{fig:error_propagation_impact_two_packetlosses}, can be detected. In the first one, $\xi_{f_{a},p_{f_{a}}}<\xi_{f_{b},p_{f_{b}}}$. In the second one, $\xi_{f_{a},p_{f_{a}}}\geq\xi_{f_{b},p_{f_{b}}}$. The area that is considered affected by the loss of packet $p_{f_{a}}$ is shown in orange and the effect of packet $p_{f_{b}}$ is colored in blue for the two prediction structures considered. So, the striped areas reflect, for each packet, the pixels in each frame that are affected by the loss of this very packet, considering the loss of previous ones. As can be clearly seen, there is an overlap between the potential damaged area associated to both packets, that is, the pixels that would be affected if the packets were lost singly. This means that there exists a correlation between the effects of both losses. An initial conclusion is the assertion that models that are not aware of packet interdependence cannot accurately capture the outcome from losing more than one interrelated data packets, as they do not consider this correlation factor whatsoever. This correlation factor is, for the metric used in this approach, negative, as it must not be counted more than once. This contrasts with what happens when using PSNR or other metrics, as stated in the work by Liang et al.~\cite{2003-liang} and others.

So, considering the proposed metric, overlapped areas must be considered only once. Hence, the number of impaired pixels that a packet loss really adds to the total depends on the previous losses in the bitstream within the same window of observance, as packets are sent according to the frame decoding order. Furthermore, in the specific case where the losses take place in the
same frame, this is consistent with the loss of synchronization
that takes place. Hence, returning to Fig.~\ref{fig:error_propagation_impact_two_packetlosses}, the number of impacted pixels finally added to the total by each lost packet 
is colored accordingly, as above-mentioned. In the first case (Fig.~\ref{fig:error_propagation_impact_IPP_two_packetlosses_a} and Fig.~\ref{fig:error_propagation_impact_hier_two_packetlosses_a}),
the contribution of the loss of the packet colored in blue to the
overall distortion introduced in the sequence, initially estimated
assuming that the frame it belongs to is independent, is reduced by
the simultaneous loss of a packet in a reference frame (colored in
orange). In the second one (Fig.~\ref{fig:error_propagation_impact_IPP_two_packetlosses_b} and Fig.~\ref{fig:error_propagation_impact_hier_two_packetlosses_b}),
the contribution of the loss of the packet colored in blue to the
overall distortion is reduced from the initially estimated one to
zero if the packet colored in orange is also lost. This makes perfect
sense, since the overall distortion that is introduced as a result
of all the losses remains the same regardless of this packet being
correctly received or lost.

Finally, we present in Fig.~\ref{fig:error_propagation_impact_three_packetlosses} another example where three packets are lost. Let packet $p_{f_{a}}$ belonging to frame $f_{a}$ be the first of the three lost packets in the transmitted packet stream, packet $p_{f_{b}}$ in frame $f_{b}$ be the second one, and packet $p_{f_{c}}$ in frame $f_{c}$ be the third. The isolated loss of the first of them would cause an impaired area of $\xi_{f_{a},p_{f_{a}}}$ percent of the pixels of frame $f_{a}$ and the frames depending on it, the isolated loss of the second one would change the value of $\xi_{f_{b},p_{f_{b}}}$ percent of the pixels of frame $f_{b}$ and of the frames relying on it, and the isolated loss of the third one would change the value of $\xi_{f_{c},p_{f_{c}}}$ percent of the pixels of frame $f_{c}$ and of the frames depending on it. In the example presented in the figure, we have assume that $\xi_{f_{a},p_{f_{a}}}<\xi_{f_{b},p_{f_{b}}}$, $\xi_{f_{a},p_{f_{a}}}<\xi_{f_{c},p_{f_{c}}}$ and $\xi_{f_{b},p_{f_{b}}}>\xi_{f_{c},p_{f_{c}}}$. In accordance with the above description, we have used colors black, orange and blue to identify the impaired area that is attributed to packets $p_{f_{a}}$, $p_{f_{b}}$ and $p_{f_{c}}$, respectively. It can be seen that the contribution of the third packet is zero, as in the previous example.

\begin{figure*}[tp]
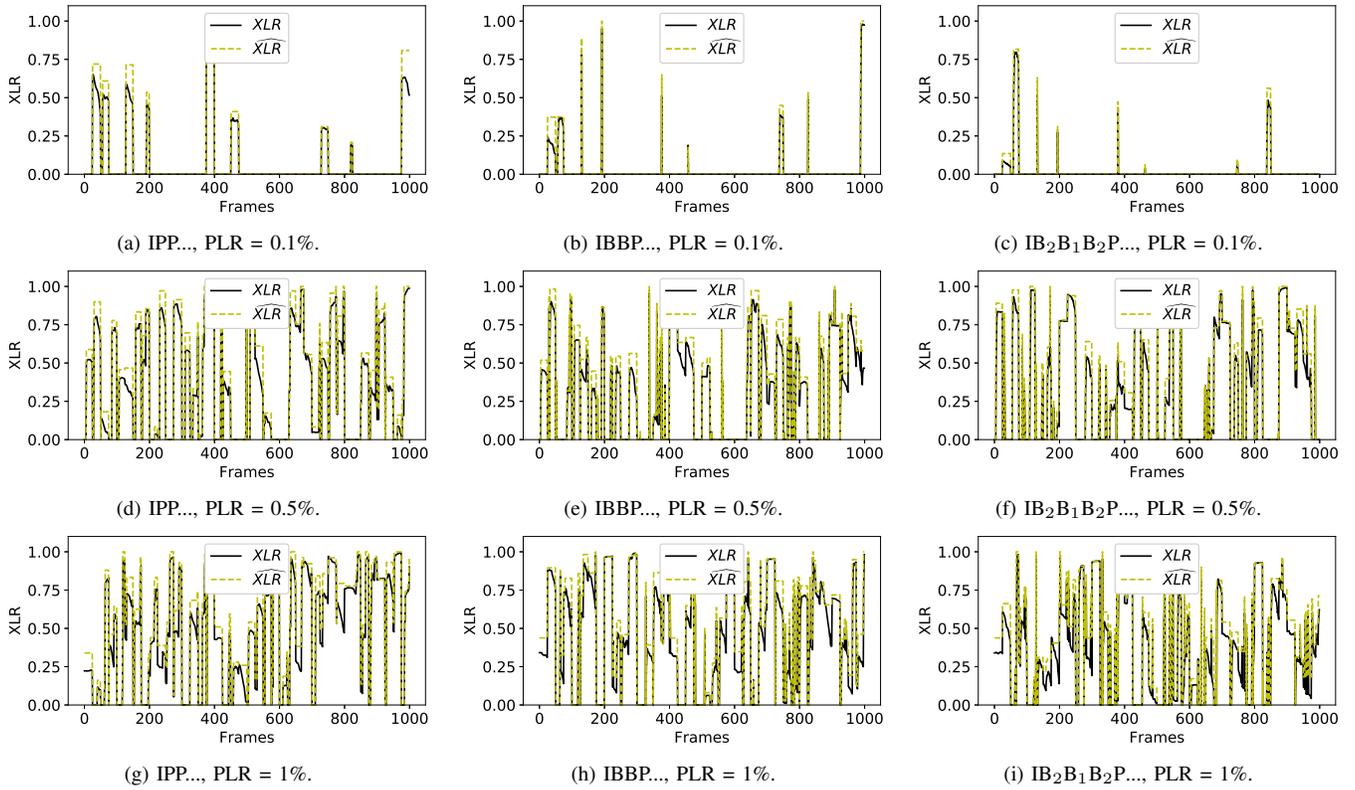

\centering
\subfloat[\label{fig:real_vs_estimated_xlr_seeking_IPP_plr_1}IPP..., PLR~=~0.1\%.]{
\includegraphics[width=0.32\textwidth]{{{Figures/pixel_loss_rate_seeking_1920x1080_IPP_GOP25_IDR_VBR_8Mbps.ts_trace_ch_G_plr_0.001_abl_2p.txt}}}
}
\subfloat[\label{fig:real_vs_estimated_xlr_seeking_IBBP_plr_1}IBBP...,
PLR~=~0.1\%.]{
\includegraphics[width=0.32\textwidth]{{{Figures/pixel_loss_rate_seeking_1920x1080_IB2P_GOP25_IDR_VBR_8Mbps.ts_trace_ch_G_plr_0.001_abl_2p.txt}}}
}
\subfloat[\label{fig:real_vs_estimated_xlr_seeking_IB2B1B2P_plr_1}$\text{IB}_{2}\text{B}_{1}\text{B}_{2}\text{P}$...,
PLR~=~0.1\%.]{
\includegraphics[width=0.32\textwidth]{{{Figures/pixel_loss_rate_seeking_1920x1080_IB3Phier_GOP25_IDR_VBR_8Mbps.ts_trace_ch_G_plr_0.001_abl_2p.txt}}}
}\\\vspace{-.3cm}
\subfloat[\label{fig:real_vs_estimated_xlr_seeking_IPP_plr_5}IPP..., PLR~=~0.5\%.]{
\includegraphics[width=0.32\textwidth]{{{Figures/pixel_loss_rate_seeking_1920x1080_IPP_GOP25_IDR_VBR_8Mbps.ts_trace_ch_G_plr_0.005_abl_2p.txt}}}
}
\subfloat[\label{fig:real_vs_estimated_xlr_seeking_IBBP_plr_5}IBBP...,
PLR~=~0.5\%.]{
\includegraphics[width=0.32\textwidth]{{{Figures/pixel_loss_rate_seeking_1920x1080_IB2P_GOP25_IDR_VBR_8Mbps.ts_trace_ch_G_plr_0.005_abl_2p.txt}}}
}
\subfloat[\label{fig:real_vs_estimated_xlr_seeking_IB2B1B2P_plr_5}$\text{IB}_{2}\text{B}_{1}\text{B}_{2}\text{P}$...,
PLR~=~0.5\%.]{
\includegraphics[width=0.32\textwidth]{{{Figures/pixel_loss_rate_seeking_1920x1080_IB3Phier_GOP25_IDR_VBR_8Mbps.ts_trace_ch_G_plr_0.005_abl_2p.txt}}}
}\\\vspace{-.3cm}
\subfloat[\label{fig:real_vs_estimated_xlr_seeking_IPP_plr_10}IPP..., PLR~=~1\%.]{
\includegraphics[width=0.32\textwidth]{{{Figures/pixel_loss_rate_seeking_1920x1080_IPP_GOP25_IDR_VBR_8Mbps.ts_trace_ch_G_plr_0.01_abl_2p.txt}}}
}
\subfloat[\label{fig:real_vs_estimated_xlr_seeking_IBBP_plr_10}IBBP...,
PLR~=~1\%.]{
\includegraphics[width=0.32\textwidth]{{{Figures/pixel_loss_rate_seeking_1920x1080_IB2P_GOP25_IDR_VBR_8Mbps.ts_trace_ch_G_plr_0.01_abl_2p.txt}}}
}
\subfloat[\label{fig:real_vs_estimated_xlr_seeking_IB2B1B2P_plr_10}$\text{IB}_{2}\text{B}_{1}\text{B}_{2}\text{P}$...,
PLR~=~1\%.]{
\includegraphics[width=0.32\textwidth]{{{Figures/pixel_loss_rate_seeking_1920x1080_IB3Phier_GOP25_IDR_VBR_8Mbps.ts_trace_ch_G_plr_0.01_abl_2p.txt}}}
}
\caption{\label{fig:real_vs_estimated_xlr_seeking}Comparison, frame by
frame, for the sequence 'seeking',
of the proposed FR and NR metrics. The actual computed XLR values
obtained as a result of the FR metric are depicted in black and the
estimated XLR values resulting from applying the proposed NR metric
are shown in yellow.}
\end{figure*}

\begin{figure*}[tp]
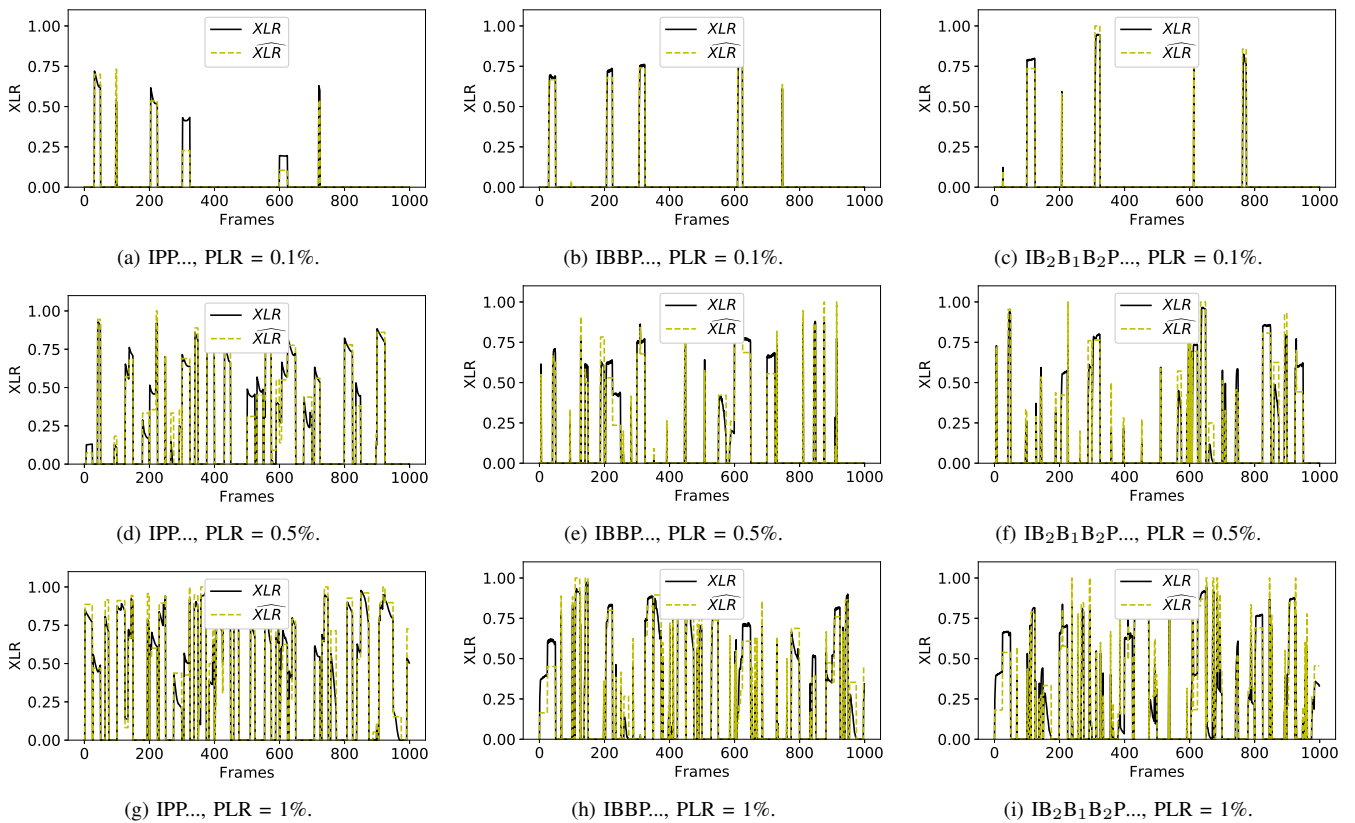

\centering
\subfloat[\label{fig:real_vs_estimated_xlr_elfuente2_IPP_plr_1}IPP..., PLR~=~0.1\%.]{
\includegraphics[width=0.32\textwidth]{{{Figures/pixel_loss_rate_elfuente2_1920x1080_IPP_GOP25_IDR_VBR_8Mbps.ts_trace_ch_G_plr_0.001_abl_2p.txt}}}
}
\subfloat[\label{fig:real_vs_estimated_xlr_elfuente2_IBBP_plr_1}IBBP...,
PLR~=~0.1\%.]{
\includegraphics[width=0.32\textwidth]{{{Figures/pixel_loss_rate_elfuente2_1920x1080_IB2P_GOP25_IDR_VBR_8Mbps.ts_trace_ch_G_plr_0.001_abl_2p.txt}}}
}
\subfloat[\label{fig:real_vs_estimated_xlr_elfuente2_IB2B1B2P_plr_1}$\text{IB}_{2}\text{B}_{1}\text{B}_{2}\text{P}$...,
PLR~=~0.1\%.]{
\includegraphics[width=0.32\textwidth]{{{Figures/pixel_loss_rate_elfuente2_1920x1080_IB3Phier_GOP25_IDR_VBR_8Mbps.ts_trace_ch_G_plr_0.001_abl_2p.txt}}}
}\\\vspace{-.3cm}
\subfloat[\label{fig:real_vs_estimated_xlr_elfuente2_IPP_plr_5}IPP..., PLR~=~0.5\%.]{
\includegraphics[width=0.32\textwidth]{{{Figures/pixel_loss_rate_elfuente2_1920x1080_IPP_GOP25_IDR_VBR_8Mbps.ts_trace_ch_G_plr_0.005_abl_2p.txt}}}
}
\subfloat[\label{fig:real_vs_estimated_xlr_elfuente2_IBBP_plr_5}IBBP...,
PLR~=~0.5\%.]{
\includegraphics[width=0.32\textwidth]{{{Figures/pixel_loss_rate_elfuente2_1920x1080_IB2P_GOP25_IDR_VBR_8Mbps.ts_trace_ch_G_plr_0.005_abl_2p.txt}}}
}
\subfloat[\label{fig:real_vs_estimated_xlr_elfuente2_IB2B1B2P_plr_5}$\text{IB}_{2}\text{B}_{1}\text{B}_{2}\text{P}$...,
PLR~=~0.5\%.]{
\includegraphics[width=0.32\textwidth]{{{Figures/pixel_loss_rate_elfuente2_1920x1080_IB3Phier_GOP25_IDR_VBR_8Mbps.ts_trace_ch_G_plr_0.005_abl_2p.txt}}}
}\\\vspace{-.3cm}
\subfloat[\label{fig:real_vs_estimated_xlr_elfuente2_IPP_plr_10}IPP..., PLR~=~1\%.]{
\includegraphics[width=0.32\textwidth]{{{Figures/pixel_loss_rate_elfuente2_1920x1080_IPP_GOP25_IDR_VBR_8Mbps.ts_trace_ch_G_plr_0.01_abl_2p.txt}}}
}
\subfloat[\label{fig:real_vs_estimated_xlr_elfuente2_IBBP_plr_10}IBBP...,
PLR~=~1\%.]{
\includegraphics[width=0.32\textwidth]{{{Figures/pixel_loss_rate_elfuente2_1920x1080_IB2P_GOP25_IDR_VBR_8Mbps.ts_trace_ch_G_plr_0.01_abl_2p.txt}}}
}
\subfloat[\label{fig:real_vs_estimated_xlr_elfuente2_IB2B1B2P_plr_10}$\text{IB}_{2}\text{B}_{1}\text{B}_{2}\text{P}$...,
PLR~=~1\%.]{
\includegraphics[width=0.32\textwidth]{{{Figures/pixel_loss_rate_elfuente2_1920x1080_IB3Phier_GOP25_IDR_VBR_8Mbps.ts_trace_ch_G_plr_0.01_abl_2p.txt}}}
}
\caption{\label{fig:real_vs_estimated_xlr_elfuente2}Comparison, frame by frame,
for the sequence 'elfuente2', of
the proposed FR and NR metrics. The actual computed XLR values obtained
as a result of the FR metric are depicted in black and the estimated
XLR values resulting from applying the proposed NR metric are shown
in yellow.}
\end{figure*}

\subsection{Formalization}
Based on the analysis and formalization carried out before, we here introduce a lightweight, yet highly accurate, procedure to estimate the distortion.

As discussed, the impact of not receiving a packet is modulated by the effect of losing previous packets in the bitstream. The masking effect identified in the previous analysis allows to properly estimate the impaired area of every frame in the sequence.

The XLR score corresponding to frame $f$, $\text{XLR}_{f}$, can be easily estimated taking into account that packet losses accumulate impaired pixels in dependent frames. To do so, let $p_{f_{a}}$ in frame $f_{a}$, $p_{f_{b}}$ in frame $f_{b}$, ..., and $p_{f_{i}}$ in frame $f_{i}$ be the set of lost packets carrying information of $f$ or of its references. Frames $f_{x}$ and $f_{y}$, where $x,y \subset \{a,b,... , i\}$, can be the same frame or different ones. Then, the XLR on frame $f$, that is, the number of impaired pixels in this frame, would equal the greatest impaired area of all the packets. This is formalized as follows: 
\begin{equation}
\widehat{\text{XLR}}_{f}=\text{max}(\xi_{f_{a},p_{f_{a}}},\xi_{f_{b},p_{f_{b}}},...,\xi_{f_{i},p_{f_{i}}})
\end{equation}
where $\xi_{f_{a},p_{f_{a}}}$, $\xi_{f_{b},p_{f_{b}}}$, ..., $\xi_{f_{i},p_{f_{i}}}$ are calculated as in~(\ref{eq:impaired_area_frame}). Of course, if no packets carrying information of $f$ or of its references are lost, the number of impaired pixels in this frame is zero.

Note that the effect of losing synchronization is included in the expressions, as the distortion introduced by a packet that is not the first lost one of a frame is always zero.

\begin{table*}[!t]
\renewcommand{\arraystretch}{1.3}
\caption{Obtained MXLR and MSXLR Values and statistical analysis results: MAE, PCC and SROCC.}
\label{tab:results}
\centering
\begin{tabular}{|c|c|c|c|c|c|c|c|c|c|}
\hline
\multirow{2}{*}{\textbf{Sequence}} & \textbf{Prediction} & \multirow{2}{*}{\textbf{PLR}} & \textbf{Real} & \textbf{Estimated} & \textbf{Real} & \textbf{Estimated} & \multirow{2}{*}{\textbf{MAE}} & \multirow{2}{*}{\textbf{PCC}} & \multirow{2}{*}{\textbf{SROCC}}\\
 & \textbf{structure} &  & \textbf{MXLR} & \textbf{MXLR} & \textbf{MSXLR} & \textbf{MSXLR} & & &\\
\hline 
\multirow{9}{*}{seeking} & \multirow{3}{*}{IPP...} & 0.1\% & 0.092 & 0.110 & 0.125 & 0.137 & 0.019 & 0.988 & 0.999\\
\cline{3-10} 
 &  & 0.5\% & 0.359 & 0.417 & 0.464 & 0.508 & 0.058 & 0.989 & 0.990\\
\cline{3-10} 
 &  & 1\% & 0.485 & 0.551 & 0.601 & 0.648 & 0.066 & 0.986 & 0.983\\
\cline{2-10} 
 & \multirow{3}{*}{IBBP...} & 0.1\% & 0.029 & 0.037 & 0.044 & 0.051 & 0.007 & 0.977 & 1.000\\
\cline{3-10} 
 &  & 0.5\% & 0.314 & 0.367 & 0.415 & 0.451 & 0.053 & 0.985 & 0.988\\
\cline{3-10} 
 &  & 1\% & 0.450 & 0.515 & 0.574 & 0.622 & 0.067 & 0.980 & 0.978\\
\cline{2-10} 
 & \multirow{3}{*}{$\text{IB}_{2}\text{B}_{1}\text{B}_{2}\text{P}$...} & 0.1\% & 0.020 & 0.025 & 0.031 & 0.036 & 0.005 & 0.989 & 1.000\\
\cline{3-10} 
 &  & 0.5\% & 0.321 & 0.367 & 0.399 & 0.431 & 0.046 & 0.985 & 0.991\\
\cline{3-10} 
 &  & 1\% & 0.365 & 0.434 & 0.502 & 0.559 & 0.070 & 0.972 & 0.982\\
\hline 
\multirow{9}{*}{elfuente2} & \multirow{3}{*}{IPP...} & 0.1\% & 0.042 & 0.037 & 0.062 & 0.056 & 0.009 & 0.969 & 1.000\\
\cline{3-10} 
 &  & 0.5\% & 0.246 & 0.256 & 0.316 & 0.326 & 0.036 & 0.982 & 0.994\\
\cline{3-10} 
 &  & 1\% & 0.389 & 0.417 & 0.464 & 0.486 & 0.050 & 0.980 & 0.981\\
\cline{2-10} 
 & \multirow{3}{*}{IBBP...} & 0.1\% & 0.052 & 0.053 & 0.062 & 0.062 & 0.003 & 0.995 & 1.000\\
\cline{3-10} 
 &  & 0.5\% & 0.145 & 0.137 & 0.182 & 0.176 & 0.022 & 0.982 & 0.997\\
\cline{3-10} 
 &  & 1\% & 0.255 & 0.262 & 0.319 & 0.325 & 0.049 & 0.965 & 0.984\\
\cline{2-10} 
 & \multirow{3}{*}{$\text{IB}_{2}\text{B}_{1}\text{B}_{2}\text{P}$...} & 0.1\% & 0.044 & 0.046 & 0.050 & 0.051 & 0.005 & 0.989 & 0.999\\
\cline{3-10} 
 &  & 0.5\% & 0.143 & 0.155 & 0.181 & 0.191 & 0.031 & 0.966 & 0.995\\
\cline{3-10} 
 &  & 1\% & 0.247 & 0.252 & 0.322 & 0.331 & 0.053 & 0.959 & 0.986\\
\hline 
\multirow{9}{*}{tennis} & \multirow{3}{*}{IPP...} & 0.1\% & 0.071 & 0.086 & 0.105 & 0.120 & 0.021 & 0.958 & 0.995\\
\cline{3-10} 
 &  & 0.5\% & 0.391 & 0.422 & 0.499 & 0.520 & 0.058 & 0.962 & 0.968\\
\cline{3-10} 
 &  & 1\% & 0.523 & 0.538 & 0.630 & 0.643 & 0.057 & 0.976 & 0.961\\
\cline{2-10} 
 & \multirow{3}{*}{IBBP...} & 0.1\% & 0.051 & 0.053 & 0.069 & 0.072 & 0.009 & 0.968 & 0.999\\
\cline{3-10} 
 &  & 0.5\% & 0.337 & 0.340 & 0.413 & 0.418 & 0.035 & 0.982 & 0.988\\
\cline{3-10} 
 &  & 1\% & 0.457 & 0.468 & 0.525 & 0.537 & 0.040 & 0.986 & 0.975\\
\cline{2-10} 
 & \multirow{3}{*}{$\text{IB}_{2}\text{B}_{1}\text{B}_{2}\text{P}$...} & 0.1\% & 0.019 & 0.021 & 0.035 & 0.037 & 0.003 & 0.989 & 1.000\\
\cline{3-10} 
 &  & 0.5\% & 0.250 & 0.255 & 0.333 & 0.341 & 0.037 & 0.971 & 0.989\\
\cline{3-10} 
 &  & 1\% & 0.492 & 0.491 & 0.577 & 0.577 & 0.042 & 0.984 & 0.975\\
\hline 
\multirow{9}{*}{birdscage} & \multirow{3}{*}{IPP...} & 0.1\% & 0.019 & 0.043 & 0.046 & 0.067 & 0.024 & 0.963 & 1.000\\
\cline{3-10} 
 &  & 0.5\% & 0.234 & 0.382 & 0.356 & 0.461 & 0.148 & 0.944 & 0.974\\
\cline{3-10} 
 &  & 1\% & 0.278 & 0.500 & 0.431 & 0.580 & 0.222 & 0.958 & 0.931\\
\cline{2-10} 
 & \multirow{3}{*}{IBBP...} & 0.1\% & 0.021 & 0.032 & 0.034 & 0.045 & 0.011 & 0.951 & 1.000\\
\cline{3-10} 
 &  & 0.5\% & 0.241 & 0.390 & 0.360 & 0.466 & 0.149 & 0.944 & 0.956\\
\cline{3-10} 
 &  & 1\% & 0.379 & 0.545 & 0.537 & 0.655 & 0.166 & 0.947 & 0.940\\
\cline{2-10} 
 & \multirow{3}{*}{$\text{IB}_{2}\text{B}_{1}\text{B}_{2}\text{P}$...} & 0.1\% & 0.022 & 0.043 & 0.038 & 0.055 & 0.021 & 0.973 & 1.000\\
\cline{3-10} 
 &  & 0.5\% & 0.197 & 0.307 & 0.323 & 0.408 & 0.110 & 0.965 & 0.982\\
\cline{3-10} 
 &  & 1\% & 0.431 & 0.603 & 0.593 & 0.706 & 0.173 & 0.950 & 0.932\\
\hline 
\multicolumn{3}{|c|}{\textbf{PCC of}} & \multicolumn{2}{c|}{\multirow{2}{*}{0.958}} & \multicolumn{2}{c|}{\multirow{2}{*}{0.987}} & \multicolumn{3}{c|}{\multirow{2}{*}{-}}\\
\multicolumn{3}{|c|}{\textbf{MXLR and MSXLR}} &  \multicolumn{2}{c|}{\multirow{2}{*}{}} &  \multicolumn{2}{c|}{\multirow{2}{*}{}} & \multicolumn{3}{c|}{\multirow{2}{*}{}}\\
\hline
\end{tabular}
\end{table*}

\subsection{Model evaluation}

The performance of the proposed model is evaluated by examining how
the estimated XLR correlates with the real value for a number of bitstreams
that are the result of simulated unreliable transmissions. This comparison
is carried out by computing/estimating the XLR frame by frame and
the overall MXLR/MSXLR. The real XLR is the result of applying~\eqref{eq:xlr},
that is, of obtaining the number of pixels whose value actually differs between
the original, non-corrupted stream and the one that has been affected
by a series of packet losses due to an unreliable transmission.

The four Full HD sequences used in Section~\ref{sec:noref_est_XLR} ('seeking', 'elfuente2', 'tennis' and 'birdsincage'~\cite{2016-netflix}) are used for the set of experiments. In particular, we have used the first 100 frames of each sequence. All sequences were H.264/AVC encoded using a prediction structure of 25 frames at 8~Mbps and 25~fps following different patterns. Specifically, we have employed the structures already shown in the examples before, i.e., IPP... and $\text{IB}_{2}\text{B}_{1}\text{B}_{2}\text{P}$..., and we have added a third one: IBBP... (all in display order). As indicated before, regarding the second one, the B-frames referred to as $\text{B}_{1}$ are used as reference for other B-frames and the ones referred to as $\text{B}_{2}$ are not, following a hierarchical temporal prediction structure. Considering the last one, B-frames are not used as references for other frames whatsoever. Additionally, in all the cases, in the same way as before, dependent frames may use intracoding, if it is more efficient. All I-frames are IDR frames to make all structures closed, all P-frames have as reference only the previous I- or P-frame, and all B-frames have as references the immediately previous and following reference frames (either I-, P- or $\text{B}_{1}$-frames). This is done to restrict the variability of the problem in terms of path for error propagation, and so more easily validate the proposed model. Of course, nevertheless, conclusions can be directly extrapolated to more commonly used open prediction structures and frames with multiple references. Finally, to obtain more significant results, for each combination of source and prediction structure, we have concatenated the resulting encoded video ten times, thus obtaining encoded sequences of 1000 frames.

Additionally, our metric is completely independent of the pattern in which packets are lost during transmission, as it considers separately the impact of each of them regardless of if it was lost singly or together with other packets in the same error burst. So, we have simulated the unreliable transmission of the encoded sequences through a simplified Gilbert channel~\cite{2011-diaz}, i.e. a two-state Markov model where the 'Good' state always means successful packet delivery and the 'Bad' state always means that the packet is lost. The reason to use this particular model to simulate the communications channel is that it is a widely used approach which, despite its simplicity, suffices for providing an adequate sample of the situations that might be encountered that affect the application of the estimated XLR, namely combinations of single and burst packet losses in one or several contiguous video frames. Two channels with the same average burst length (2 packets) but different overall packet loss rate (0.1\%, 0.5\% and 1\%) were employed.

We include in the graphs in Fig.~\ref{fig:real_vs_estimated_xlr_seeking} and Fig.~\ref{fig:real_vs_estimated_xlr_elfuente2} the results frame by frame of the real and estimated XLR values, that is, of the FR and NR methods, respectively, for two of the sequences: 'seeking' and 'elfuente2'. The results of the rest of the tested sequences present a very similar behavior and are not depicted due to lack of space. The horizontal axis indicates the index of the frame in the sequence and the vertical axis, the calculated/estimated XLR. The actual XLR values are presented in black and the estimated values in yellow. The estimated values match quite well the actual values, although the error is slightly overestimated. Table~\ref{tab:results} presents the computed MXLR and MSXLR values for all the sequences, as well as the Mean Absolute Error, the frame-to-frame Pearson Correlation Coefficient (PCC), and the Spearman's Rank Correlation Coefficient (SROCC) between estimated and computed XLR. Aggregated XLR values (either MXLR or MSXLR) show an extremely strong correlation between the estimated and the actual value (PCC of 0.958 and 0.987, respectively), meaning that the bitstream-based NR metric is a reliable representation of the pixel-based FR one. Besides, the frame-to-frame correlation is very high as well, which shows that single-frame XLR can also be derived from the bistream in a reliable way. This means that any other pooling mechanism of XLR beyond mean or mean of square roots can be used, and the bitstream estimation will still be correct.

\section{Discussion}
\label{sec:discussion}
The usage of XLR as a KQI measurement offers interesting possibilities both for network planning and monitoring.

In the case of network planning, it builds a direct relationship between the statistical packet error pattern of the communication network and the actual effect in perceived quality (the pixel loss rate), which only depends on the coding structure of the video (i.e. GOP structure and average size, in bytes, of the different frames along the stream). Therefore, only by knowing the coding structure, it should be possible to predict the average XLR given some channel model or, reversely, to determine the maximum allowed packet error rate in the channel for a target XLR value. This will allow a more efficient allocation of resources, especially when adjusting the constrains in Ultra-Reliable Low-Latency Communication (URLCC) slices, where reliability is always achieved at the cost of reducing the usable throughput of a given Radio Access Network (RAN) slot~\cite{2018-popovski}.

In the context of QoE monitoring, XLR is a very lightweight measurement that can be implemented by any system that has access to the application layer. For the typical use cases of low-latency video communication (e.g. videoconferencing, V2X communication...), they would be the end peers, but also some network elements such as Selective Forwarding Units (SFUs)~\cite{2018-andre}, or event software-defined network functions that implement this kind of features~\cite{2019-kirmizioglu}. In all those cases, implementing XLR would not significantly impact the performance of the systems, while providing better QoE capabilities (in terms of monitoring or management) than just relying on packet loss statistics. In addition, the communication between service providers, network providers and clients for XLR reporting would not require any extra channels, since it would use the same ones as the rest of implemented KPIs and KQIs.

A particular difficulty for XLR implementation would be the presence of encrypted video, which is today ubiquitous for privacy and security reasons. For low-latency applications, where collaboration between the application and the network can be strongly beneficial for both parts, this difficulty could be circumvented by traditional techniques implemented on the client side, such as selective encryption~\cite{2003-zeng} or the addition of syntactic information in RTP header extensions~\cite{2011_perez_b}.

Going forward, XLR could even be used for adaptive streaming when the transport layer is QUIC. QUIC has been developed as a replacement of TCP transport for HTTP connections~\cite{2017-langley}. It includes end-to-end encryption to prevent 'ossification', i.e. that routing elements can understand the packets and therefore perform any kind of QoE management. However, it could be possible to implement an extension where the header information is encrypted as well and used by the peers to decide on whether to request for retransmissions or \textit{keep going}. As one of the main advantages of QUIC is its reduction of rebuffering rate in adaptive streaming video, implementing XLR analysis on the client could allow for reducing buffering rate even more at the cost of losing some pixels. The good point of this approach is that it would keep all the decisions on the end device, therefore complying with QUIC design requirements.

\section{Conclusion}
\label{sec:conclusion}
We have proposed piXel Loss Rate (XLR) as a new
Key Quality Indicator (KQI) for video QoE management in IP networks.
Unlike plain network-based measures like Packet Loss Rate, XLR captures
the effect of network errors in the received video. We have shown
that the linear-dimension version of XLR, Mean Square root of piXel
Loss Rate (MSXLR), correlates with subjective MOS in a way that is
comparable to state-of-the art Full-Reference metrics.

XLR is truly 'objective' in the sense that it is well-defined
and can be exactly computed for any transmission error, regardless
the specific codec implementation. In that aspect, it is similar to
PSNR or SSIM as non-parametrized universal metrics that can be used
to describe content impairment directly. However, it only depends
on the video coding structure, but not on the video content itself.

Based on this property, we have described a technique that allows
estimating XLR for each individual frame using only a lightweight
analysis of the coded bitstream, which only requires knowing the position
of the loss within a coded frame, and the coding structure of the
video (type of each frame within the prediction structure). This information can be
obtained with very low computational power, as well as no latency,
therefore making it suitable for QoE monitoring in IP networks. Besides,
it can be used to predict the effect of transmission errors given
a channel model and a statistical distribution of frame sizes, thus
making it useful as well as network planning tool.

We have shown that this No-Reference technique can estimate the actual
XLR of each frame with great accuracy, thus making the actual XLR
value and its No-Reference estimation basically equivalent. Consequently,
we can affirm that XLR is a good predictor of video QoE under network
packet losses, which can be used both as network monitoring and planning
tool. Therefore, it is a good KQI for low-latency and high-throughput
scenarios, increasingly common in modern IP networks.

\bibliographystyle{IEEEtran}
\bibliography{refs_tnsm2019}

\begin{IEEEbiography}[{\includegraphics[width=1in,height=1.25in,clip,keepaspectratio]{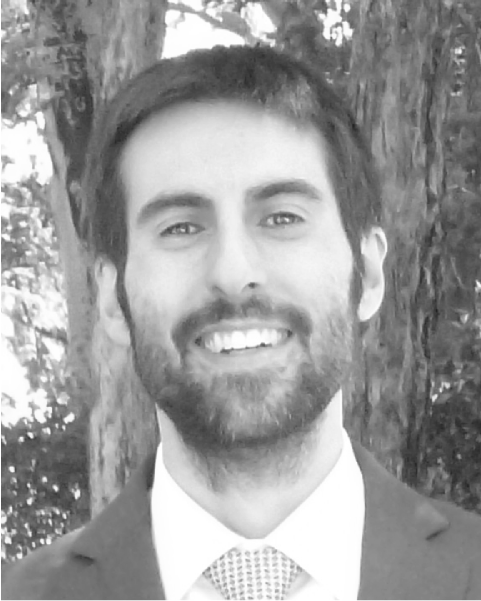}}]{C\'esar D\'iaz}
 received the Telecommunication Engineering degree (integrated BSc-MS)
in 2007 and the Ph.D. degree in Telecommunication Engineering in 2017,
both from the Universidad Polit\'ecnica de Madrid (UPM), Madrid, Spain.
Since 2008 he has been a member of the Image Processing Group of the
UPM, where he has been actively involved in Spanish and European projects.
His research interests are in the area of multimedia delivery and immersive communications.
\end{IEEEbiography}

%\vskip 0pt plus -1fil

\begin{IEEEbiography}[{\includegraphics[width=1in,height=1.25in,clip,keepaspectratio]{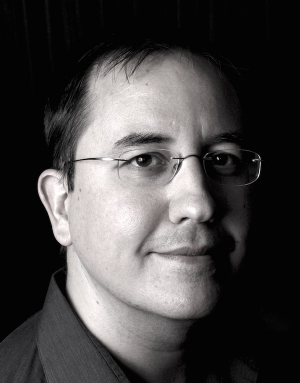}}]{Pablo P\'erez}
 received the Telecommunication Engineering degree (integrated BSc-MS)
in 2004 and the Ph.D. degree in Telecommunication Engineering in 2013
(Doctoral Graduation Award), both from Universidad Polit\'ecnica de
Madrid (UPM), Madrid, Spain. From 2004 to 2006 he was a Research Engineer in the Digital Platforms Television in Telef\'onica I+D and, from 2006
to 2017, he has worked in the R\&D department of the video business
unit in Alcatel-Lucent (later acquired by Nokia), serving as technical
lead of several video delivery products. Since 2017, he is Senior
Researcher in the Distributed Reality Solutions department at Nokia
Bell Labs. His research interests include multimedia quality of experience, video transport networks, and immersive communications.
\end{IEEEbiography}

%\vskip 0pt plus -1fil

\begin{IEEEbiography}[{\includegraphics[width=1in,height=1.25in,clip,keepaspectratio]{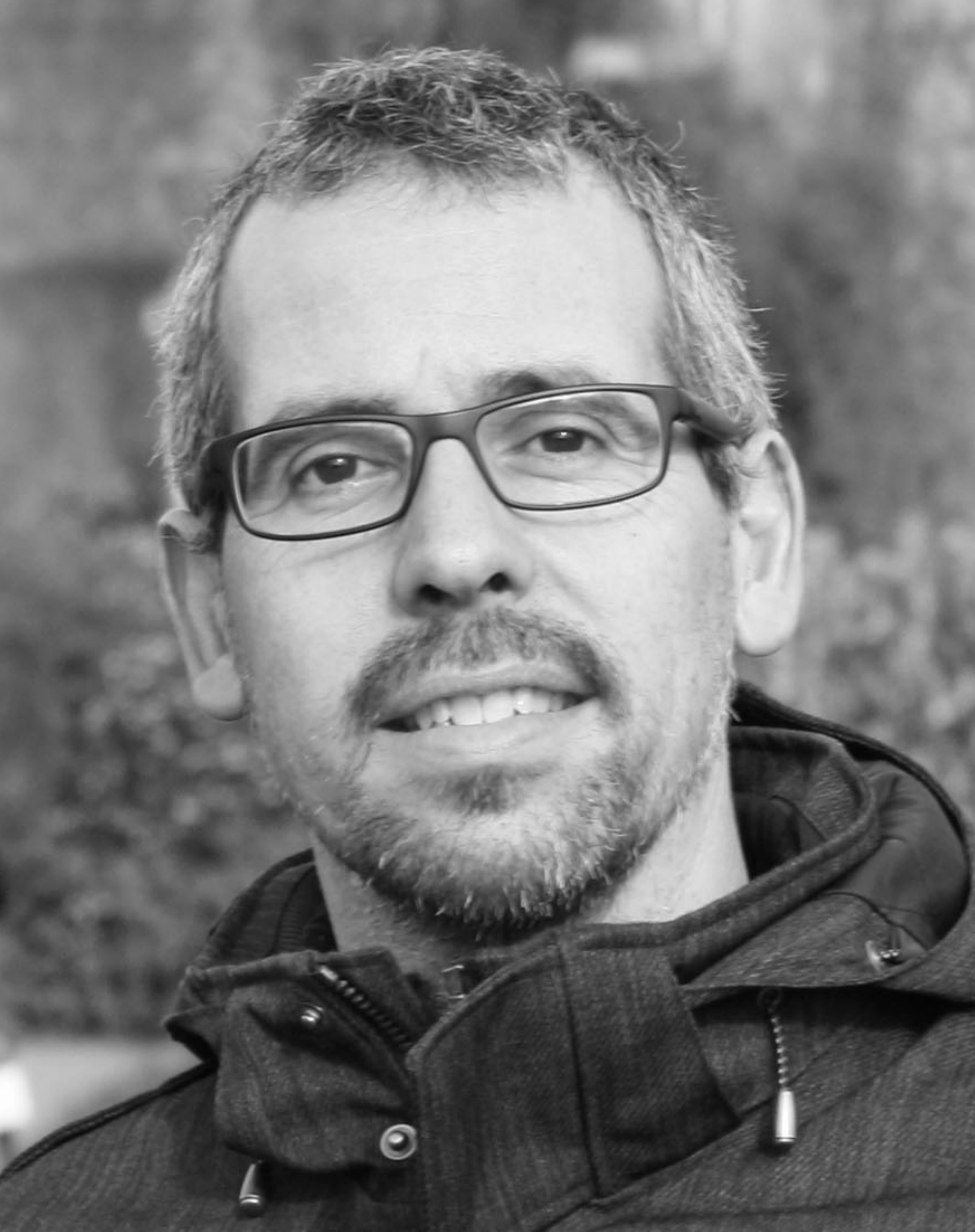}}]{Juli\'an Cabrera}
received the Telecommunication Engineering and Ph.D. degrees in telecommunication from the Universidad Polit\'ecnica de Madrid (UPM), in 1996 and 2003, respectively. Since 1996, he is a member of the Image Processing Group, UPM. Since 2001, he has been a member of the faculty of the UPM, and since 2003, he has been an Associate Professor of signal theory and communications. Current research interests cover several topics related to audio-visual communications, advance video coding, 3D and Multiview scenarios, depth estimation with special focus on deep learning approaches, video subjective quality assessment for Multiview and VR360 video, and optimization of adaptive streaming techniques.
\end{IEEEbiography}

%\vskip 0pt plus -1fil

\begin{IEEEbiography}[{\includegraphics[width=1in,height=1.25in,clip,keepaspectratio]{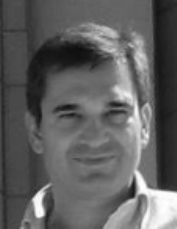}}]{Jaime J. Ruiz}
 received the Telecommunication Engineering degree (integrated BSc-MS)
in 1993 from the Universidad Polit\'ecnica de Madrid (UPM), Madrid,
Spain, and the Ph.D. degree in Industrial Engineering in 2012 from
the Universidad Nacional de Educaci\'on a Distancia (UNED), Madrid,
Spain. He holds also a Master in Marketing and Commercial Management
from ESIC, Madrid, Spain. He is now a member of the Distributed Reality
Solutions team in Bell Labs Madrid, where he is investigating on video
technology. He has leaded several R\&D projects, and has a deep experience
and knowledge on video performance solutions involving distributed
storage and streaming technology. His research interest include performance
video solutions, low latency implementations, high performant configurations,
and massive video storage and streaming, mainly focused on 360 Video
and Immersive VR video.
\end{IEEEbiography}

%\vskip 0pt plus -1fil

\begin{IEEEbiography}[{\includegraphics[width=1in,height=1.25in,clip,keepaspectratio]{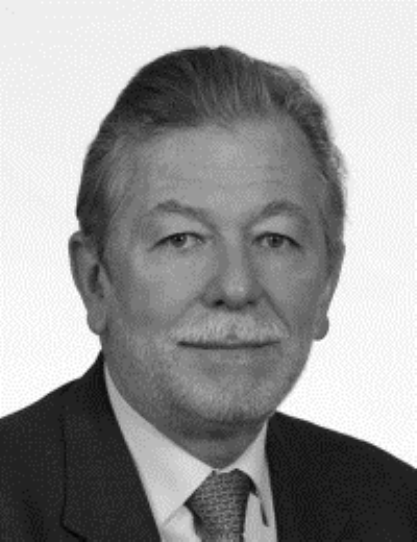}}]{Narciso Garc\'ia}
 received the Ingeniero de Telecomunicaci\'on degree (five years engineering
program) in 1976 (Spanish National Graduation Award) and the Doctor
Ingeniero de Telecomunicaci\'on degree (PhD in Communications) in 1983
(Doctoral Graduation Award), both from the Universidad Polit\'ecnica
de Madrid (UPM), Madrid, Spain. Since 1977, he has been a member of
the faculty of the UPM, where he is currently a Professor of Signal
Theory and Communications. He leads the Grupo de Tratamiento de Im\'agenes
(Image Processing Group), UPM. He has been actively involved in Spanish
and European research projects, also serving as an evaluator, a reviewer,
an auditor, and an observer of several research and development programs
of the European Union. He was a co-writer of the EBU proposal, base
of the ITU standard for digital transmission of TV at 34--45 Mb/s
(ITU-T J.81). He was Area Coordinator of the Spanish Evaluation Agency
(ANEP) from 1990 to 1992 and he has been the General Coordinator of
the Spanish Commission for the Evaluation of the Research Activity
(CNEAI) from 2011 to 2014. He has been the Vice-Rector for International
Relations of the Universidad Polit\'ecnica de Madrid from 2014 to 2016.
His current research interests include digital image and video compression
and computer vision. Dr. Garc\'ia received the Junior and Senior Research
Awards of the Universidad Polit\'ecnica de Madrid in 1987 and 1994,
respectively.
\end{IEEEbiography}

\end{document}